\documentclass{aa}

\usepackage{graphicx,psfig,amssymb,verbatim,natbib}

\bibpunct{(}{)}{;}{a}{}{,}

\newcommand\ciii{\hbox{C$\rm III]~\lambda1909$~\AA}}
\newcommand\se{Sect.~}
\newcommand\si{0156--252~}
\newcommand\sii{0406--244~}
\newcommand\siii{0828+193~}
\newcommand\siv{2036--254~}
\newcommand\sv{2048--272~}
\newcommand\chandra{{\it Chandra}~}

\begin{document}      
              
\title{
On the X-ray emission of $z\sim2$ radio galaxies: IC scattering of the CMB \& no evidence for 
fully-formed potential wells
}

\subtitle{}

   \author{R. A. Overzier\inst{1}
          \and D. E. Harris\inst{2}
          \and C. L. Carilli\inst{3} 
          \and L. Pentericci\inst{4}
          \and H. J. A. R\"ottgering\inst{1}
          \and G. K. Miley\inst{1}}

   \offprints{R.~A. Overzier}

   \institute{Leiden Observatory, University of Leiden, P.O. Box 9513, 2300 RA Leiden, The Netherlands\\
              \email{overzier@strw.leidenuniv.nl}
\and
Smithsonian Astronomical Observatory, Harvard-Smithsonian Center for Astrophysics, 60 Garden Street, Cambridge, MA 02138, USA\\
\and 
National Radio Astronomy Observatory, New Mexico Array Operations Center (VLA, VLBA), P.O. Box O, 1003 Lopezville Road, Socorro, NM 87801, USA\\
\and
Dipartimento di Fisica, Universit\`a degli Studi Roma Tre, Rome, Italy\\
}

   \date{Received <date> / Accepted <date>}

   \abstract{We present the results of 20 ksec \chandra observations for each of 5 radio galaxies in the redshift range $2.0<z<2.6$. 
The goals were to (i) study the nature of their non-thermal X-ray emission, (ii) investigate  
the presence and amount of hot gas, and (iii) look for active galactic nuclei (AGN) overdensities in 
fields around high redshift radio galaxies. 
For 4 of the 5 targets we detect unresolved X-ray 
components coincident with the radio nuclei. From spectral analysis of one of the cores and comparison to the empirical radio to X-ray luminosity ratio ($L_R/L_X$) correlation for AGN, we find that the cores 
are underluminous in the X-rays indicating that obscuring material ($n(\hbox{H\rm I})\sim10^{22}$ cm$^{-2}$) may be surrounding the nuclei. 

We detect X-ray emission coincident with the radio hotspots or lobes in 4 of the 5 targets. 
This extended emission can be explained by the Inverse-Compton (IC) scattering of photons that make up the 
cosmic microwave background (CMB). The magnetic field strengths of $\sim100-200$ $\mu G$ that we derive agree 
with the equipartition magnetic field strengths. The relative ease with which the lobe X-ray emission is
detected is a consequence of the $(1+z)^4$ increase in the energy density of the CMB. For one of the lobes,  
the X-ray emission could also be produced by a reservoir of hot, shocked gas. An HST image of the region around this radio 
component shows bright optical emission reminiscent of a bow-shock.

By co-adding the 5 fields we created a deep, 100 ksec exposure to search for diffuse X-ray emission from thermal 
intra-cluster gas. We detect no diffuse emission and derive upper limits of $\sim10^{44}$ erg s$^{-1}$, thereby ruling out a virialized structure of cluster-size scale at $z\sim2$. 

The average number of soft X-ray sources in the field surrounding the radio sources is consistent with the number density 
of AGN in the \chandra Deep Fields, with only one of the fields showing a marginally statistically significant 
factor 2 excess of sources with 
$f_{0.5-2keV}>3\times10^{-15}$ erg s$^{-1}$ cm$^{-2}$. Analysis of the angular distribution of the field sources shows no evidence for large-scale 
structure associated with the radio galaxies, as was observed in the case of PKS 1138--262 
by \citet{laura02}. 

   \keywords{galaxies: high-redshift -- galaxies: active -- X-rays: galaxies: clusters -- X-rays: general}
   }

\authorrunning{R.~A. Overzier et al.}
\titlerunning{\chandra observations of $z\sim2$ radio galaxies}
   \maketitle


\section{Introduction}
\label{sec:intro}

Radio galaxies can be used to trace the formation and evolution of the most massive galaxies 
known at high redshift. They usually have continuum morphologies suggestive of the merging of $L_*$  
systems \citep[e.g.][]{laura99,laura01}, and are surrounded by large reservoirs of line emitting gas 
\citep[e.g.][]{ojik97} that are comparable in size to cD galaxy envelopes. 
High redshift radio galaxy (HzRG) fields targeted by broad and narrow band imaging and 
spectroscopy have been found to locate 'protoclusters'. They have large excesses of Ly$\alpha$ and/or H$\alpha$ 
emitters, Lyman break galaxies and extremely red objects \citep[e.g.][]{laura00,kurk00,kurk03,venemans02,miley04}. 
Thus, HzRGs act as beacons to the progenitors of present-day galaxy clusters 
\citep[see also e.g.][]{windhorst98,ivison00,brand03,smail03,stevens03}.
Finding these distant (i.e. $z\gtrsim2$) protoclusters is important for constraining models of structure formation and 
cosmology.

Radio continuum polarimetric observations of HzRGs have revealed large rotation 
measures (RM) and large RM gradients. At $z\gtrsim2$ the typical RM observed is several hundreds 
rad m$^{-2}$, but extreme values of $>1000$ rad m$^{-2}$ are not uncommon \citep{carilli97,athreya98,pentericci00}. 
It is believed that RM arises from 
'Faraday screens' in our galaxy (i.e. the interstellar medium, ISM), in intervening galaxies and clusters,  
or near the source itself. 
At low redshifts, most radio galaxies have relatively small RM values that can be attributed to the 
local ISM \citep[][and references therein]{simard80,leahy87}. 
However, RM values observed towards HzRGs are significantly larger, and are believed to be due to 
Faraday screening of extra-galactic origin somewhere along the line of sight. 
\citet{athreya98} show that the chance superpositions of distant radio sources and intervening systems such as 
foreground Abell clusters and damped Ly$\alpha$ absorbers are too small to explain the fraction of high RMs observed. 
Likewise, the expected RMs from such intervening systems are too small to account for the high values seen 
towards distant radio sources. Hence, the large RM values may occur in a medium close to the source, although 
it is unclear whether they are caused by extended cluster-sized media or smaller gas distributions on scales comparable 
to the radio sources \citep[e.g.][]{carilli02}. Some radio galaxies at low redshift have extremely large values 
of RM that are caused by Faraday rotation in so-called 'cooling flow' cluster atmospheres 
\citep[e.g. 3C 295;][see \citet{carilli02b} for a review on cluster magnetic fields]{perley91}. There is evidence that the strength of cooling flows is correlated 
with RM, indicating that RM is an indicator of the fields in the intracluster medium rather than 
the radio cocoon \citep[][but see \citet{rudnick03}]{taylor94,eilek02,ensslin03}. 




X-ray observations are also a powerful tool to find deep, gravitational potential wells in the early Universe.  
Diffuse X-ray emission caused by thermal bremsstrahlung from a hot ICM, is evidence for a bound system in 
which the gas is in dynamical equilibrium with the galaxies and the cold dark matter. Deep X-ray surveys have 
uncovered significant numbers of rich galaxy clusters out to $z\sim1.2$ \citep[e.g.][]{rosati99,stanford01,rosati04}. 
Remarkably, the diffuse ICM in these distant objects traces the galaxy distribution, 
and their surface brightness and temperature profiles are similar to those of lower redshift clusters, indicating 
that clusters formed very early in the history of the Universe. The study of thermal emission from HzRGs is 
important because of their suggested linkage to cluster formation. However, if an ICM 
is present in such structures at $z>1.5$, the extreme cosmological surface brightness dimming makes their 
detection very difficult. 

So far only a handful of distant radio sources has been studied with {\it Chandra}. 
\citet{fabian03} performed a very deep (200 ks) study of the radio galaxy 3C 294 at $z=1.79$. 
They observed a 100 kpc region of diffuse emission bounded by sharp edges. Although some thermal component 
from the ICM could not be ruled out, most of the emission was ascribed to Inverse Compton (IC) 
scattering of the cosmic microwave background (CMB) by an older population of electrons 
tracing out an hourglass-shaped region around the radio source. \citet{belsole04} found weak cluster luminosities and 
IC scattering among a sample of three bright HzRGs. 
\citet{carilli02} found diffuse X-ray emission around the radio galaxy PKS 1138--262 at $z=2.16$, 
which is believed to be the forming, massive galaxy at the center of a protocluster \citep{laura00,kurk00,kurk03}. 
However, the extended emission is seen only along the radio axis, and is therefore believed to be 
associated with shocked material inside the radio source. 
The upper limit that \citet{carilli02} derive for the X-ray luminosity of the ICM  
($\sim40$\% of the luminosity of the Cygnus A cluster, see e.g. \citet{smith02}) is not unexpectedly low  
because there has not been enough time since the big bang for these structures at $z\sim2$ 
to form a sufficiently deep potential well. 
\citet{scharf03} detected extended X-ray emission around 4C 41.17 at $z=3.8$  
roughly following the radio morphology. They conclude that the X-ray emission arises from 'Inverse-Compton 
scattering of far-infrared photons from a relativistic electron population probably associated with past and current 
activity from the central object', in addition to a lesser contribution from the up-scattering of CMB photons.      

\begin{table*}[t]
\caption[]{\label{tab:sample}Log of observations. The last column lists the average galactic $\hbox{H\rm I}$ column density in the direction of each source as given by \citet{dickey90}.}
\begin{center}
\begin{tabular}[t]{ccccccc}
\hline
\hline
Source & R.A. (J2000) & Decl. (J2000) & $z$ & Dates of observation & Exposure time (s) & $n(\hbox{H\rm I})$ ($\times10^{20}$ cm$^{-2}$)\\
\hline
\si   & 01$^h$58$^m$33.5$^s$ & $-24\degr$59$\arcmin$32.26$\arcsec$ &    2.09 &   2002-11-09 & 11477 &1.34\\
&&&&  2002-12-20  &8401\\
\sii  & 04$^h$08$^m$51.5$^s$ & $-24\degr$18$\arcmin$16.8$\arcsec$ &     2.44 &  2002-12-07&   20179&3.30\\
\siii & 08$^h$30$^m$53.4$^s$ &  $+19\degr$13$\arcmin$15.7$\arcsec$ &    2.57 &  2002-11-05 &  19212&3.72\\
\siv  & 20$^h$39$^m$24.5$^s$ & $-25\degr$14$\arcmin$30.4$\arcsec$ &     2.00 &   2002-09-01&   19660&4.92\\
\sv   & 20$^h$51$^m$03.6$^s$ & $-27\degr$03$\arcmin$02.1$\arcsec$ &     2.06 &  2002-11-11 &   19555&6.29\\

\hline
\end{tabular}
\end{center}
\end{table*}

It is clear that the nature of the X-ray emission mechanisms operating in radio galaxies is diverse. 
Better sample statistics are required to enlarge our current understanding of e.g. 
the dominant X-ray emission mechanisms, the role of magnetic fields on galaxy and cluster-size scales, 
the origin of the hot gas observed in X-ray luminous clusters, and large-scale 
structure associated with HzRGs. Here we present observations with the \chandra X-ray observatory 
of five additional radio galaxies at $z\sim2$. 
The sources were selected from a compendium of $z>2$ HzRGs, focusing on the lower end of 
the redshift distribution in order to minimize surface brightness dimming effects, while still being 
at high enough redshift to allow for cosmological evolution effects in source properties. Sources were further 
selected on the presence of one or more of the various characteristics:  
distorted radio morphology,  
large rotation measure ($\ge1000$ rad m$^{-2}$), evidence for interaction 
between Ly$\alpha$ and radio structure, and extended Ly$\alpha$ absorption features.    

Throughout this paper we assume 
$H_0=70$ km s$^{-1}$, $\Omega_M=0.3$, and $\Omega_\Lambda=0.7$.
The resulting scale factors range from 8.0 to 8.4 kpc arcsec$^{-1}$ from $z=2.6$ to $z=2$. 
We will use power-law spectra defined as $S_\nu\propto\nu^{-\alpha}$, where $\alpha$ is the spectral index.  


\section{Observations} 
\label{sec:sample}

\subsection{Observations and data reduction} 

Each of the 5 sources was observed for 20 ks with the back-illuminated ACIS-S3 chip on {\it Chandra}, 
using the standard 3.2 s readout timed exposure mode and the faint telemetry format. 
Sources were positioned at approximately 20\arcsec\ away from the aimpoint. 
The data were processed and analysed in June 2003 using the \chandra data analysis package 
CIAO 2.3\footnote{http://cxc.harvard.edu/ciao/} together with CALDB 2.22. We applied the 'ACISABS' script that corrects the \chandra auxiliary response files 
for a continuous degredation in the ACIS quantum efficiency. We did not apply the 'tgain' correction that has 
become standard practice only in the most recent release of 
CIAO to correct for the drift in the effective detector gain due to an increase in the charge transfer 
inefficiency over time. However, for ACIS--S3 the resulting effect on the measured photon energies was of the 
order of only $\sim0.3$\% near the end of 2002 when the data were taken.

A log of the observations is given in Table 1. \si was observed on 2 separate dates, because the 
initial observation was interrupted after 11.5 ks by a high radiation shutdown of the system due to an 
excess of solar wind particles. To ease the analysis of these observations, the two exposures were merged into a single event file by 'reprojecting' the events.
A lightcurve analysis of the other sources indicates no further periods of significant fluctuations in the count rates of source and large, source-free background regions.

Radio images at 5 and 8 GHz where obtained by \citet{carilli97} and \citet{laura00} using 
the VLA in A configuration. The noise is 50 $\mu$Jy/beam at 5 Ghz and 25 $\mu$Jy/beam at 8 GHz. The resolution of the observations 
is 0.43\arcsec\ for the 5 GHz maps and 0.23\arcsec\ for the 8 GHz maps. Analysis of the radio data was performed using the Astronomical Image Processing System (AIPS) and the Multichannel Image Reconstruction, Image Analysis and Display (MIRIAD) software.   

We will also make use of observations obtained with the Wide-Field Planetary Camera 2 (WFPC2) on HST. 
\si and \sv were observed as part of Program 8183 (PI: Miley), and an image of \sii was retrieved from the 
HST archive (program 8338, PI: Lehnert). \si was observed for 4800 s and \sv for 7200 s, both through filter $F555W$. 
This filter has a central wavelength of 5443 \AA. \sii was observed for 2000 s through filter $F675W$ which has a 
central wavelength of 6718 \AA. Measurements 
from these images were made using the conversion from counts to flux density as given in the PHOTFLAM header keyword of each image. 
To get to the flux density $S_\nu$ (erg s$^{-1}$ cm$^{-2}$ Hz$^{-1}$) we calculate $S_\nu=(\lambda^2/c)\times(counts/t_{exp})\times PHOTFLAM$ \citep[see][for a description of procedures]{laura99}.


\subsection{Image registration}

To recover the inherent resolution of the \chandra mirror/detector system, we 
removed the pixel randomization added in the pipeline processing and regridded our maps to 1/10 native ACIS pixel size. 
After applying a suitable smoothing function, we were able to find the position of the X-ray core to an accuracy of 
better than 0.1\arcsec\ provided the core had sufficient signal to noise (S/N). We then shifted the images to align 
the X-ray core with the radio core. 
The typical shift needed was one half of an original \chandra pixel in right ascension and/or 
declination, consistent with the known astrometric accuracy.  
In the case of \sv no shift was applied, because the nucleus was not detected in either the radio or the X-rays. 

\subsection{Analysis}

\begin{figure*}[p]
\begin{minipage}[t]{\textwidth}
\begin{center}
\mbox{
\psfig{figure=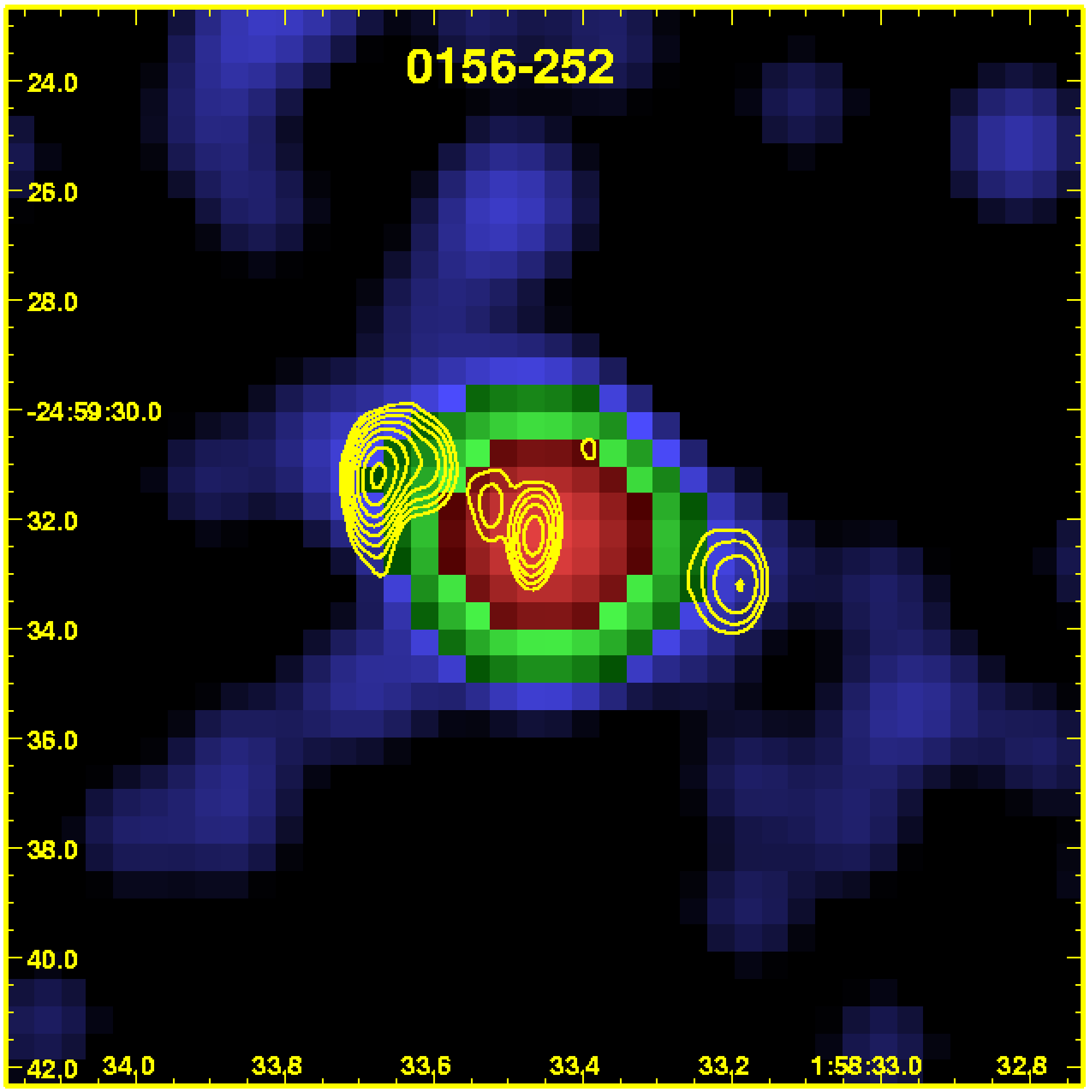,width=0.4\textwidth,clip=boundingbox}
\hspace{2cm}
\psfig{figure=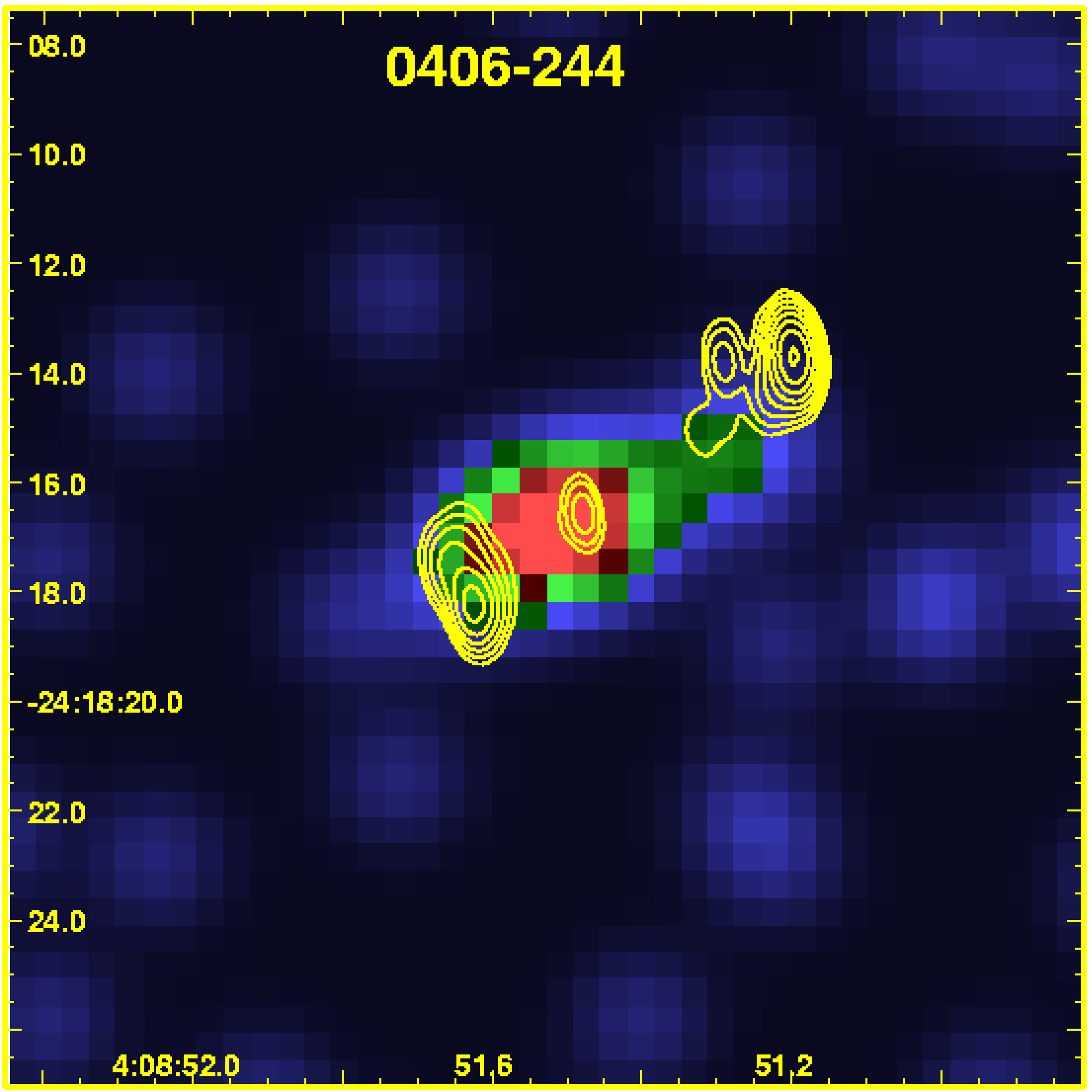,width=0.4\textwidth,clip=boundingbox}}
\end{center}
\begin{center}
\mbox{
\psfig{figure=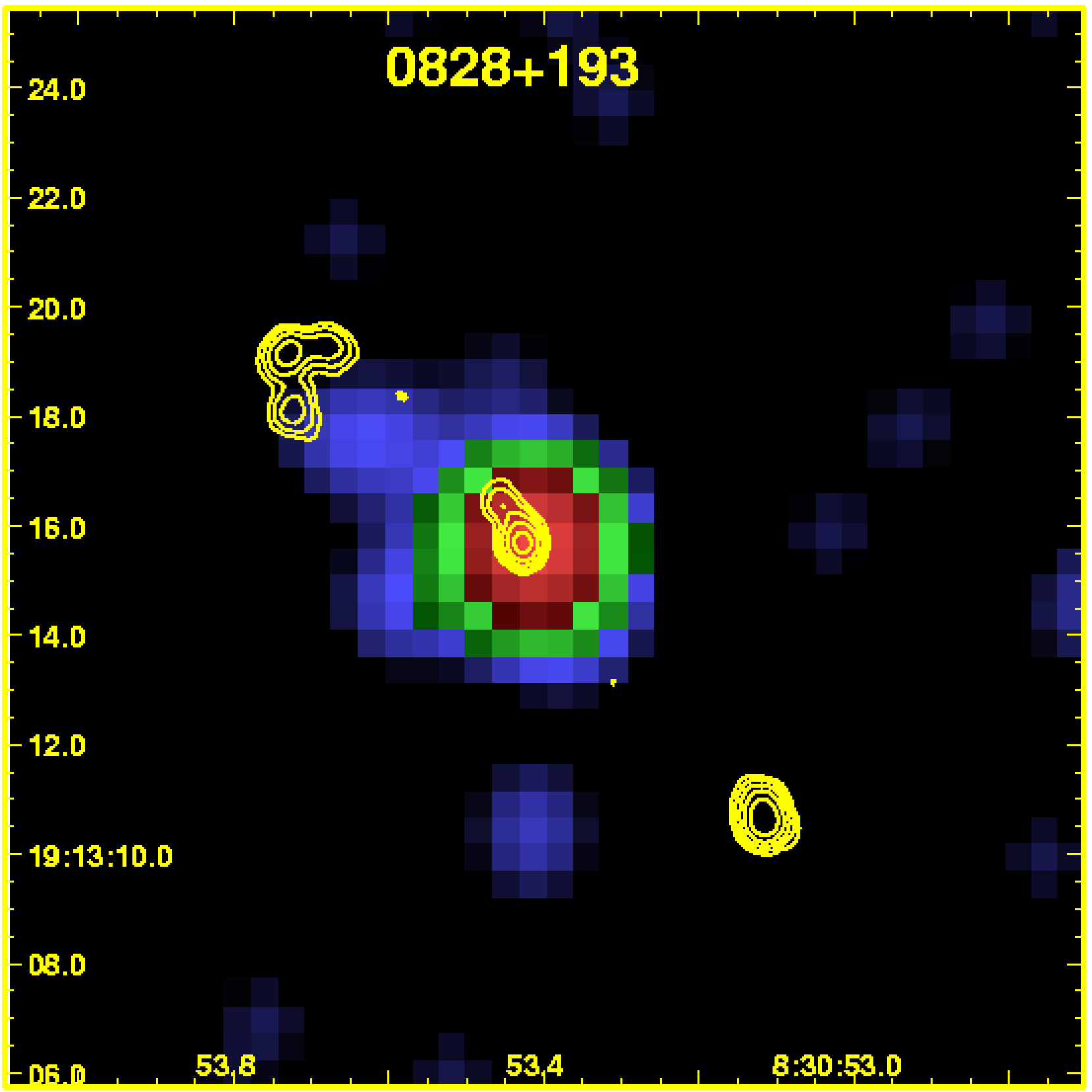,width=0.4\textwidth,clip=boundingbox}
\hspace{2cm}
\psfig{figure=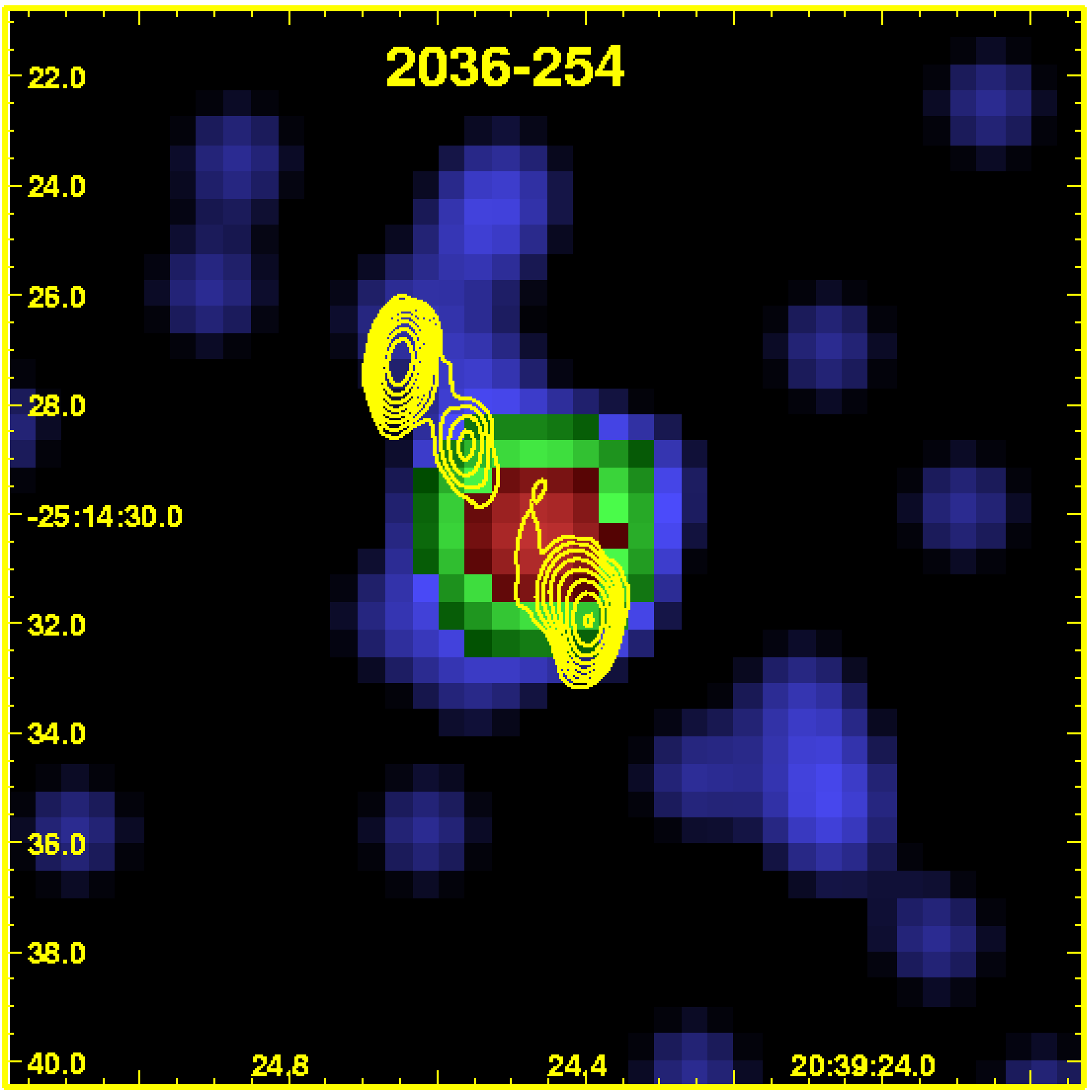,width=0.4\textwidth,clip=boundingbox}}
\end{center}

\begin{center}
\mbox{
\psfig{figure=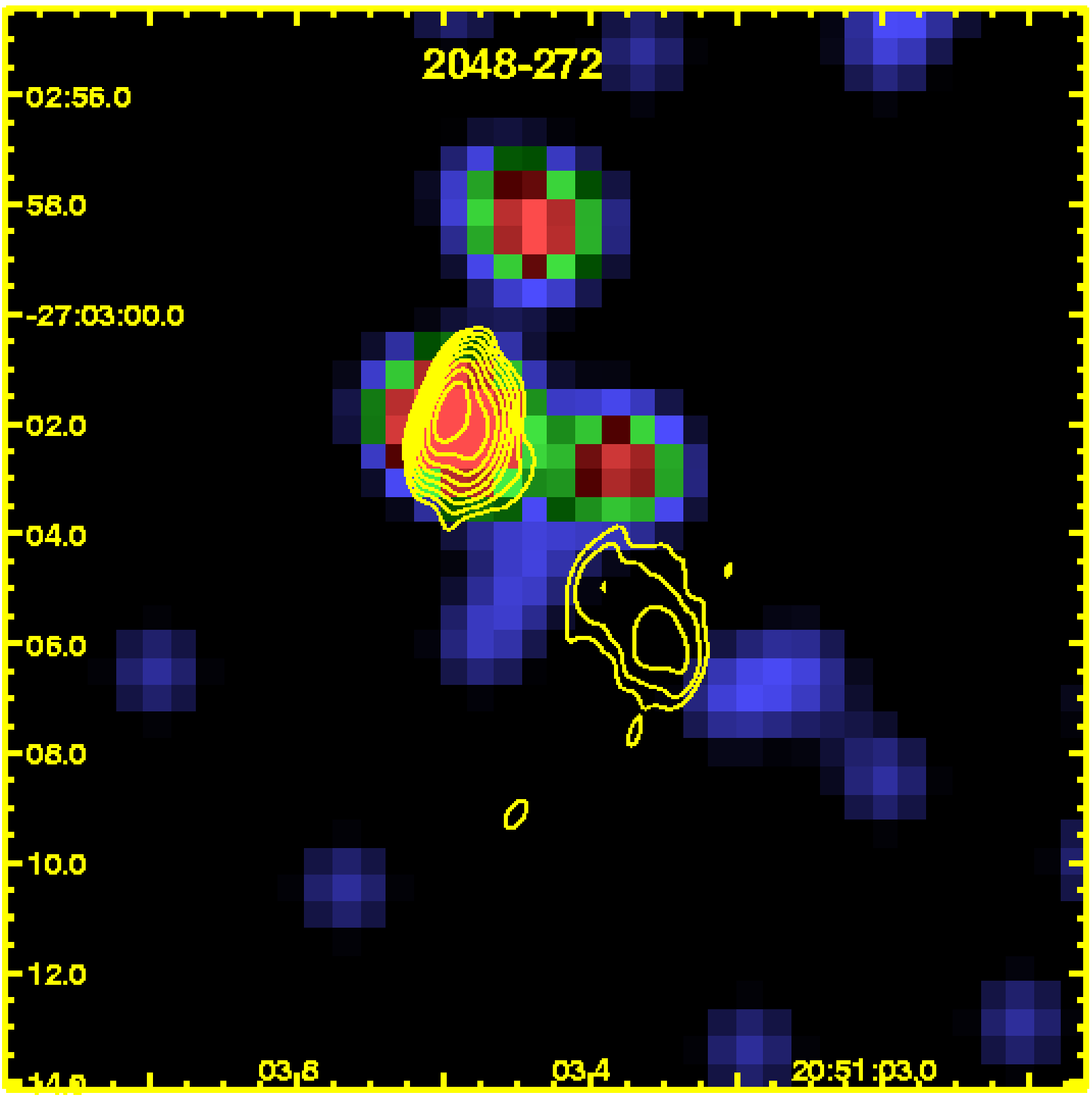,width=0.4\textwidth,clip=boundingbox}}
\end{center}
\end{minipage}
\caption{\label{fig:sources}\chandra images of 0156--252, 0406--244, 0828+193, 2036--254 and 2048--272, showing the 0.2--6 keV X-ray images in 
colorscales, and the 4.7 GHz VLA radio contours in yellow \citep[from][]{carilli97,laura00}. The $20\arcsec\times20\arcsec$ \chandra images have been smoothed using a Gaussian kernel of width 2\arcsec\ (FWHM). Radio contours are $1.0 \times (0.2, 0.4, 0.8, 1.6, ...)$ mJy beam$^{-1}$.}
\end{figure*}

The \chandra images are shown in Fig. \ref{fig:sources}, where yellow contours indicate the 4.7 GHz radio maps from \citet{carilli97} and \citet{pentericci00}. We measured the X-ray counts from the distinct radio 
components (e.g. cores and lobes) 
using boxed or circular extraction regions. The counts are background-subtracted by using a large, point source free,  
annular region centered on the radio source. We restrict our measurements to counts in the energy range of 0.2--6 keV 
for which the contribution of the background is minimal. We calculate the total  
net (i.e. background subtracted) counts $N=S-B$, where $S$ and $B$ are the counts in the source region $A_S$ and the 
background region $A_B$, respectively. We consider a radio component detected in the X-rays when $N$ is greater than its formal error, 
$\sigma_N=[(\sigma_S)^2+(\sigma_B)^2]^{1/2}$, where $\sigma_S=1+(S+0.75)^{1/2}$ and $\sigma_B=[1+(B+0.75)^{1/2}](A_S/A_B)$ 
from \citet{gehrels86}. For undetected components we calculate $2\sigma$ upper limits in the 0.2--6 keV band 
using the above expression for $\sigma_N$. The upper limit on flux is estimated by assuming a power-law 
spectrum with a spectral index of 0.8 using the online tool PIMMS\footnote{http://heasarc.gsfc.nasa.gov/Tools/w3pimms.html}.    

The procedure to convert counts into fluxes is as follows:  
we create exposure maps for the soft 
(0.5--2 keV) and hard (2--6 keV) 
band at nominal energies, and divide the images in each band by the corresponding exposure map to obtain fluxed images. 
When extracting the flux for a given region, we calculate weight factors defined by the ratio of the average energy of counts in that region to the nominal energy of each exposure map. The total flux in a region is then calculated by summing the weighted fluxes in the two bands. We remark that our method is slightly different compared to the usual procedure of calculating fluxes from exposure maps using a conversion factor that is based on the spectral response within the given band as well as an assumed spectral energy distribution. Our method is preferred because it allows us to measure fluxes, without having to assume a specific spectral shape.

The measurements are given in Table \ref{tab:counts}. 
Where indicated, fluxes are corrected for galactic absorption using PIMMS, assuming a power-law spectrum with a spectral index of 0.8 and taking the $\hbox{H\rm I}$ column density in the direction of each source from \citet{dickey90} indicated in Table \ref{tab:sample}. All energy ranges 
are in the observed frame, unless stated otherwise.

\section{Results}

\subsection{Source description}

\begin{table}[t]
\caption[]{\label{tab:RM}Rotation measures, RM$_{int}=RM_{obs}\times(1+z)^2$, for the Northern (N.) and Southern (S.) lobes of each source 
taken from the literature.}
\begin{center}
\begin{tabular}[t]{cccc}
\hline
\hline
\multicolumn{1}{c}{Source} & \multicolumn{2}{c}{RM$_{int}$ (rad m$^{-2}$)} & \multicolumn{1}{c}{Ref.}\\ 
 & N. lobe & S. lobe & \\
\hline
\si   & $1528\pm296$ & -- & 1\\
\sii  & $880\pm19$ & $-705\pm30$ & 1\\
\siii & -- & -- & 2\\
\siv  & $3\pm227$ & $-3321\pm429$ & 1\\
\sv   & $590$ & -- & 3\\
\hline
\multicolumn{4}{l}{1 Athreya et al. (1998)}\\
\multicolumn{4}{l}{2 Carilli et al. (1997)}\\
\multicolumn{4}{l}{3 Pentericci et al. (2000)}\\
\end{tabular}
\end{center}
\end{table}

We will briefly summarize some of the main characteristics of each source. Details on 
radio observations, ground-based imaging and spectroscopy, and Hubble Space Telescope (HST) observations 
can be found in \citet{mccarthy96}, \citet{carilli97}, \citet{laura99}, \citet{laura01}, \citet{iwamuro03} and references therein.  
Table \ref{tab:RM} summarizes the intrinsic rotation measures, RM$_{int}=RM_{obs}\times(1+z)^2$. 

\noindent $\bullet$ 0156--252 at $z=2.09$ has several characteristics reminiscent 
of the well-studied radio galaxy PKS 1138--262 \citep{carilli02,laura00}. Near-infrared continuum shows a host galaxy extended over $\sim2\arcsec$ parallel to the radio axis, and narrow-band Ly$\alpha$ shows emission line gas 
extended over the entire region encompassed by the northern and southern radio hot spots. 
The Ly$\alpha$ brightness is at a minimum at the 
position of the host galaxy and peaks at the location where there is a sharp bend in the (northern) radio jet, 
probably due to shock-induced ionization of the gas where the radio jet is deflected by denser material. The northern hot spot has an intrinsic RM of $\sim1000$ rad m$^{-2}$. \\
$\bullet$ 0406--244 at $z=2.44$ consists of two main optical/infrared continuum features 
aligned with the radio axis (see \citet{rush97} for a detailed study).
The continuum components lie embedded in a large Ly$\alpha$ halo with a spatial extent of $\sim8\arcsec$ along the radio 
axis. Both radio lobes have an intrinsic RM of $\sim1000$ rad m$^{-2}$ with opposite signs.\\  
$\bullet$ 0828+193 at $z=2.57$ is the largest radio source in our sample ($\sim100$ kpc). It has a jet-like 
feature north of the radio core. 
The northern lobe contains several hot spots with a peculiar 90 degree bend. The southern radio hotspot is unresolved. 
\citet{laura99} have identified the 
radio core with the brightest optical component of several clumps in a $\sim2\arcsec$ region seen by HST. 

No RM is given for this source in Table \ref{tab:RM}, since the fractional polarization at 8.2 GHz for both lobes 
was below the $4\sigma$ sensitivity limit of the observations \citep{carilli97}.\\    
$\bullet$ 2036--254 at $z=2.00$ is the only source in our sample for which no optical/near-infrared data are available. 
The radio core is most likely the faint, compact component close to the southern lobe. 
There is a pair of hot spots parallel to the radio axis in the north. 
The southern lobe has a RM$_{int}$ of $\sim3500$ rad m$^{-2}$.\\
 $\bullet$ 2048--272 at $z=2.06$ is among 
the $\sim30$\% of sources in the sample of \citet{carilli97} for which the core is undetected. However, a bright near-infrared object seen in between the lobes is presumed to be the galaxy hosting the radio source. 
The northern hot spot has a RM$_{int}$ of $\sim590$ rad m$^{-2}$. \\ 

\begin{table*}[t]
\caption[]{\label{tab:counts}X-ray measurements.}
\begin{center}
\begin{tabular}[t]{ccccccccc}
\hline
\hline
\multicolumn{1}{c}{Source} & $\theta^e$ &  Counts$^a$ & \multicolumn{1}{c}{Soft$^b$}& \multicolumn{1}{c}{Hard$^b$} &\multicolumn{1}{c}{$f_{0.5-2 keV}^c$} & \multicolumn{1}{c}{$f_{2-6 keV}^c$} & \multicolumn{1}{c}{log($L_{0.5-6 keV}$)}\\
 & ($\arcsec$) & (0.2--6 keV) &(0.5--2 keV)  & (2--6 keV)& \multicolumn{1}{c}{erg s$^{-1}$ cm$^{-2}$} & \multicolumn{1}{c}{erg s$^{-1}$ cm$^{-2}$} & \multicolumn{1}{c}{erg s$^{-1}$}\\
\hline
\si   & &        &                &                  &                             &                                &             \\
Total &4.2 & $275\pm18$& $200 \pm 14.2$ &  $69   \pm  8.3$ & $(3.1\pm0.2)\times10^{-14}$ & $(4.9\pm0.6)\times10^{-14}$ & 45.42       \\
Core  &1.1 & $249\pm17$& $181 \pm 13.5$ &  $62   \pm  7.9$ & $(2.8\pm0.2)\times10^{-14}$ & $(4.4\pm0.6)\times10^{-14}$ & 45.37       \\
NE    &1.6 & $7\pm4$& $4.9 \pm 2.2$  &  $1.0  \pm  1.0$ & $(7.4\pm3.4)\times10^{-16}$ & $(3.9\pm4.7)\times10^{-16}$    & 43.56       \\
SW    &1.2 & $4\pm3$& $1.9 \pm 1.4$  &  $1.9  \pm  1.4$ & $(4.6\pm3.4)\times10^{-16}$ & $(1.6\pm1.2)\times10^{-15}$    & $43.82$     \\
\hline                                                                                                                                   
\sii  & &        &                &                  &                             &                                &            \\
Total &4.8 & $21\pm6$       & $18  \pm 4.4$  &  $0.5  \pm  1.4$ & $(2.9\pm0.7)\times10^{-15}$ & $(3.6\pm9.3)\times10^{-16}$ & 44.19  \\
Core  &0.9 & $10\pm4$       & $8.0 \pm 2.8$  &  $2.0  \pm  1.4$ & $(1.4\pm0.5)\times10^{-15}$ & $(1.3\pm0.9)\times10^{-15}$ & 44.11  \\
SE    &1.8 & $9\pm4$       & $6.9 \pm 2.6$  &  $1.0  \pm  1.0$ & $(1.0\pm0.4)\times10^{-15}$ & $(5.2\pm6.5)\times10^{-16}$ & 43.88   \\
NW    &2.0 & ...$^d$       & \multicolumn{2}{c}{...} & \multicolumn{2}{c}{$<2.4\times10^{-15}$} & $<44.05$                           \\
\hline                                                                                                                           
\siii & &        &                &                  &                             &                             &                \\
Total &7.0 & $32\pm7$     & $17  \pm 4.4$  &  $13.3 \pm  4.0$ & $(3.2\pm0.8)\times10^{-15}$ & $(1.2\pm0.4)\times10^{-14}$ & 44.91    \\
Core  &1.0 &$23\pm6$      & $12  \pm 3.5$  &  $9.9  \pm  3.2$ & $(2.1\pm0.6)\times10^{-15}$ & $(8.7\pm2.8)\times10^{-15}$ & 44.78    \\
NE    &1.4 & ...$^d$         & \multicolumn{2}{c}{...} & \multicolumn{2}{c}{$<1.9\times10^{-15}$}  & $<44.00$                        \\
SW    &1.0 & ...$^d$         & \multicolumn{2}{c}{...} & \multicolumn{2}{c}{$<1.6\times10^{-15}$}  & $<43.93$                        \\
\hline                                                                                                                           
\siv  & &        &                &                  &                             &                             &                \\
Total &4.1 &$35\pm7$        & $14  \pm 3.9$  &  $18.5 \pm  4.4$ & $(2.6\pm0.7)\times10^{-15}$ & $(1.2\pm0.3)\times10^{-14}$ & 44.65 \\
Core  &0.6 &$22\pm6$        & $7.0 \pm 2.6$  &  $15.0 \pm  3.9$ & $(1.4\pm0.5)\times10^{-15}$ & $(9.6\pm2.5)\times10^{-15}$ & 44.51  \\
NE    &1.3 &...$^d$         & \multicolumn{2}{c}{...} & \multicolumn{2}{c}{$<2.0\times10^{-15}$}  & $<43.76$                         \\
SW    &1.4 &$6\pm4$       & $5.9 \pm 2.5$  &  $0.0  \pm  0.0$ & $(1.0\pm0.4)\times10^{-15}$ & ... & 43.47    \\
\hline                                                                                                                           
\sv   & &        &                &                  &                             &                             &                \\
Total &4.3 &$8\pm4$ & $4.1 \pm 2.2$  &  $4.0  \pm  2.2$ & $(6.0\pm3.3)\times10^{-16}$ & $(2.9\pm1.6)\times10^{-15}$ & 44.05          \\
Core  &0.8 & ...$^d$     & \multicolumn{2}{c}{...} & \multicolumn{2}{c}{$<2.1\times10^{-15}$} & $<43.82$                             \\
NE    &1.7 & $6\pm4$& $2.8 \pm 1.7$  &  $1.8  \pm  1.4$ & $(4.5\pm2.8)\times10^{-16}$ & $(1.2\pm1.0)\times10^{-15}$ & 43.73          \\
SW    &1.5 & ...$^d$     &\multicolumn{2}{c}{...}  & \multicolumn{2}{c}{$<1.7\times10^{-15}$} & $<43.72$                             \\

\hline
\end{tabular}
\end{center}
\noindent
$^a$ Errors are calculated using Gehrels (1986) in the low-count regime.\\
$^b$ Errors are $\sqrt{counts}$.\\
$^c$ Observed fluxes (i.e. not corrected for galactic absorption).\\
$^d$ Undetected. We give a $2\sigma$ upper limit for the total flux at 0.2--6 keV.\\
$^e$ Circular extraction radius.\\
\end{table*}

\subsection{X-ray Cores}

\begin{figure*}[t]
\includegraphics[width=\columnwidth]{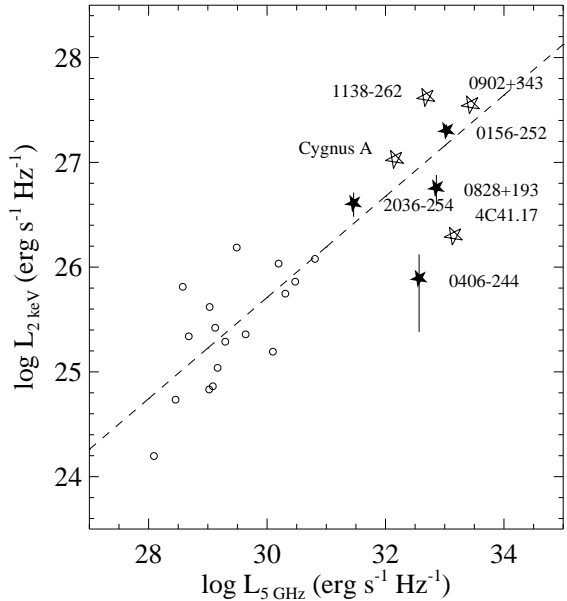}
\caption{\label{fig:lumcores}Monochromatic X-ray luminosity versus radio luminosity. The four detected cores from our sample are indicated by filled stars. Their X-ray luminosities were corrected for galactic absorption only. 
Narrow line AGN from the FIRST/ROSAT sample of 
\citet{brinkmann00} are indicated by circles. The dashed line indicates the correlation 
for radio-loud quasars of \citet{brinkmann00}. Open stars indicate several low (Cygnus A at $z=0.06$) and high redshift (1138--262 at $z=2.16$, B0902+343 at $z=3.395$, 4C41.17 at $z=3.798$) radio galaxies from the literature \citep[see][and references therein]{carilli02,scharf03,fabian02b}. The X-ray luminosities of these sources were all corrected for intrinsic absorption.}
\end{figure*}

We have detected X-ray cores in 4 of the 5 targets. No core was detected in the case of 2048--272, and we note 
that the core was also not detected in the radio maps indicating that nuclear activity may have ceased. 
The core X-ray luminosities are in the range $L_{X[0.5-6keV]}=1.3\times10^{44}-2.3\times10^{45}$ erg s$^{-1}$.   
\si has $>4\times$ the luminosity of the other cores.
We place an upper limit on the core luminosity of \sv of $7\times10^{43}$ erg s$^{-1}$. 
\citet{hardcastle99} and \citet{brinkmann00} presented a correlation between the X-ray and radio luminosity of radio-loud quasars 
from the cross-correlation of ROSAT and radio sources. It has been suggested that this correlation implies a physical 
relationship between X-ray and radio cores, which can be explained by models in which e.g. the X-ray emission originates from 
the base of the jet \citep{hardcastle99}. 
In Fig. \ref{fig:lumcores} we show objects from the \citet{brinkmann00} 
sample classified as narrow-line AGN, as well as the best-fit correlation for their subsample of radio-loud quasars, 
log$L_X=11.22+0.483\times$log$L_r$.   
We have also indicated several other $z>2$ radio galaxies (1138--262 at $z=2.16$, B0902+343 at $z=3.395$, 4C41.17 at $z=3.798$) 
and the local source Cygnus A for comparison \citep[see][and references therein]{carilli02,scharf03,fabian02b}. 

As shown in Fig. \ref{fig:lumcores}, three out of four sources in our sample agree with the $L_r/L_X$ relation if we use the observed 2--6 keV (rest-frame 6--18 keV) flux with a power-law spectrum ($\alpha=0.8$) to calculate 
the luminosity at rest-frame 2 keV (observed 0.7 keV). This result is virtually independent 
of the amount of intrinsic absorption, since it is of negligible effect at rest-frame 6--18 keV. Calculating the 2 keV luminosity from the 0.5--2 keV observed flux (rest-frame 1.5--6 keV) yields values that are significantly lower, but for 0156--252, 0828+193 and 2036--254 this can be explained by an intrinsic absorption column density of 
$n(\hbox{H\rm I})\sim10^{23}$ cm$^{-2}$. Interestingly, the core of 0406--244 seems highly underluminous in 
Fig. \ref{fig:lumcores}, a discrepancy that cannot be explained away by invoking a large amount of absorption. However, this may be a S/N issue: we have detected only 2 photons in the hard 
2--6 keV band, compared to 62, 10 and 15 for the other 3 sources. If we instead calculate the 2 keV luminosity from the 0.5--2 keV observed flux (8 photons detected) with $n(\hbox{H\rm I})=10^{23}$ cm$^{-2}$ we find log($L_{2 keV})\approx26.4$ compared to $\sim25.9$ for the earlier method, and note that its deviation from the $L_r/L_X$ relation is now comparable to that of 4C41.17.

\begin{figure}[t]
\includegraphics[width=\columnwidth]{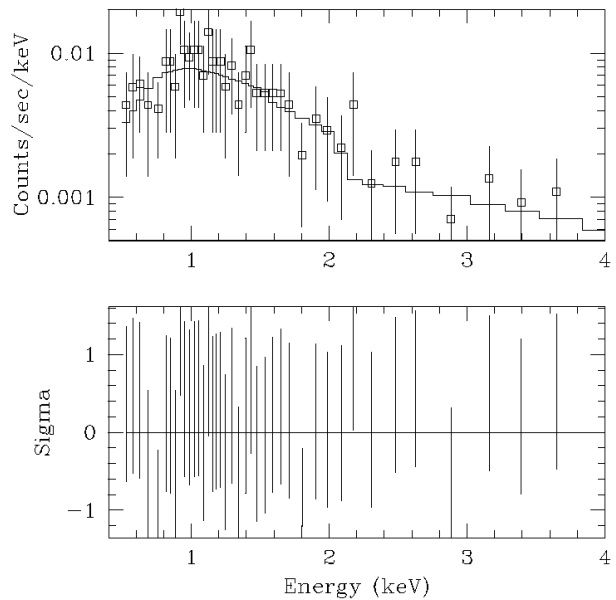}
\caption{\label{fig:spec}Spectrum of the core of 0156--252, fitted with a power-law spectrum modified by 
absorption at the source.}
\end{figure}

\si is the only source in our sample for which the core is detected with sufficient S/N to carry out a (crude) spectral analysis. The X-ray spectrum is shown in Fig. \ref{fig:spec}. Assuming a power-law spectrum with absorption at the redshift of the source, we find $\alpha_X=0.8\pm0.2$ and $n(\hbox{H\rm I})=(1.6\pm0.7)\times10^{22}$ cm$^{-2}$. The fit has a reduced $\chi^2$ of 0.34, and is indicated in Fig. \ref{fig:spec}. The 
error bars for the fit parameters correspond to a change of 1 in the reduced $\chi^2$. Repeating the fit using using the {\it Cash} statistic valid for the low counts regime we find $n(\hbox{H\rm I})=(1.5\pm0.8)\times10^{22}$ cm$^{-2}$ with $\alpha$ unchanged. The galactic $\hbox{H\rm I}$ column density towards this source is $1.34\times10^{20}$ cm$^{-2}$ (see Table 1). 
For completeness, we note that the spectrum can be fitted by a Mewe-Kaastra-Liedahl (Mekal) thermal 
plasma model with $kT=23\pm8$ keV (reduced $\chi^2$ of 0.50). 

The signal-to-noise of the 3 other cores is insufficient to make spectra. However, if we assume 
that they are a generic class we might estimate the amount of intrinsic 
absorption by stacking the data. The hardness ratio $HR\equiv(H-S)/(H+S)$ (where $H$ and $S$ are the number counts observed in 
the hard and soft bands) is $0.0\pm0.2$, consistent with a Type II AGN at $z\sim2$ with an obscured ($n(\hbox{H\rm I})\approx10^{23}$ cm$^{-2}$) power-law spectrum with a photon index of $\Gamma=1.8$ \citep{tozzi01}. 

\subsection{X-ray emission from lobes/hotspots}



We obtained positive detections of X-ray components coincident with the northeastern (NE) and southwestern (SW) radio lobes of 0156--252, the southestern (SE) lobe of 0406--244, the SW lobe of 2036--254, and the NE lobe of 2048--272. 
Below we will evaluate the likely mechanisms for these detections. Given the generally low signal to noise of all features we detected, we will focus on the X-ray emission that is identified with the radio lobes\footnote{Our radio data are inadequate to confidently identify radio features as {\it lobes} or {\it hotspots}. We adopt 
the term {\it lobes} to mean any radio emission not associated with the core.}.

\begin{figure*}[t]
\includegraphics[width=\columnwidth]{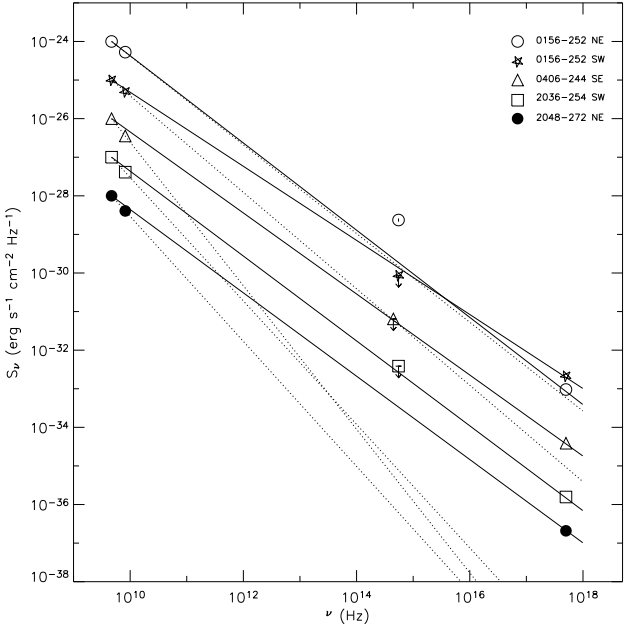}
\caption{\label{fig:sync}Synchrotron model predictions. We indicate the observed radio, optical and X-ray flux densities 
for all components that have a positive X-ray detection. Solid lines represent synchrotron spectra constructed using the 4.7 GHz radio to X-ray spectral index, $\alpha^{X}_{4.7}$. Dotted lines indicate spectra constructed using a radio spectral index, $\alpha^{8.2}_{4.7}$. 
For each component, the data and models were offset for clearer 
visibility of the results. The multiplicative factors for 0156--252 NE, 0156--252 SW, 0406--244 SE, 2036--254 SE and 2048--272 NE were 1.1, 1.4, 0.03, 0.001, 0.0003, respectively.}
\end{figure*}

\medskip
\noindent
$\bullet$ Synchrotron radiation\\
In Fig. \ref{fig:sync} we plot spectra for the lobes that have a detection in the X-ray. The spectra were constructed by (i) extrapolating the radio flux densities to X-ray frequencies using the radio spectral index 
$\alpha^{8.2}_{4.7}$, and (ii) using the 4.7 GHz to X-ray spectral index, $\alpha^{X}_{4.7}$. The results of our analysis are indicated in Table \ref{tab:synchrotron}. In the first method (indicated by dotted lines in Fig. \ref{fig:sync}), we can directly compare the observed X-ray flux density to 
the predicted synchrotron flux density based on the radio spectrum. Except for the NE lobe of 0156--252, the predicted flux densities 
are several orders of magnitude lower than that observed. This seems to rule out a synchrotron origin for the X-ray emission from these lobes. However, $S_\nu$ is a strong function of the spectral index, since we are extrapolating over $\sim10^8$ decades of frequency. Only a small offset in 
$\alpha^{8.2}_{4.7}$ can lead to a significant under- or overprediction of the X-ray flux density. The radio spectral indices used 
were measured over a relatively narrow frequency range, and may be subject to errors. We estimate that the 
typical error in $\alpha^{8.2}_{4.7}$ is $\sim0.2$, based on the comparison with a spectral index $\alpha^{8.2}_{1.4}$ determined by combining our data with data (not shown here) from the publicly available NRAO VLA Sky Survey 
(NVSS)\footnote{http://www.cv.nrao.edu/nvss/}, albeit at much lower resolution. Therefore, we attempt the second method (indicated by solid lines in Fig. \ref{fig:sync}). We slightly modify the above procedure and now estimate the radio flux density expected at 8.2 GHz, 
assuming {\it a priori} that the spectrum is synchrotron with a radio to X-ray spectral index $\alpha^{X}_{4.7}$. 
The ratio of predicted to observed 8.2 GHz flux densities is 1.0 for the NE lobe of 0156--252 and ranges from 
1.2 to 1.5 for the other lobes. This indicates that the observed X-ray flux densities are consistent with synchrotron radiation, 
provided that the error in the integrated flux densities at 8.2 GHz are $\gtrsim20$\%. Typically, errors in the integrated radio flux 
densities are of the order of $\sim5$\%. Furthermore, the extrapolation from the radio to the X-ray using the radio spectral index {\it under}-predicted the observed X-ray emission in all cases. If there were to be a large, random error in the integrated radio flux densities, we 
might have expected to see some of the spectra {\it over}-predict the X-ray flux as well. The fact that this is not observed may indicate that the typical error in the radio flux densities is indeed only a few percent.

To further test our synchrotron models, we have extracted flux densities in the optical ($\sim6000$\AA) using the 
HST/WFPC2 observations of 0156--252, 0406--244, and \sv (see \se\ref{sec:sample}). In Fig. \ref{fig:sync} 
we have indicated $1\sigma$ upper limits on the optical emission of 
the SW lobe of 0156--252, the SE lobe of \sii and the NE lobe of 2048--272. Although the flux densities in 
the optical are consistent with the synchrotron models, they are unfortunately not deep enough to rule them out. 

\begin{figure*}[t]
\includegraphics[width=\columnwidth]{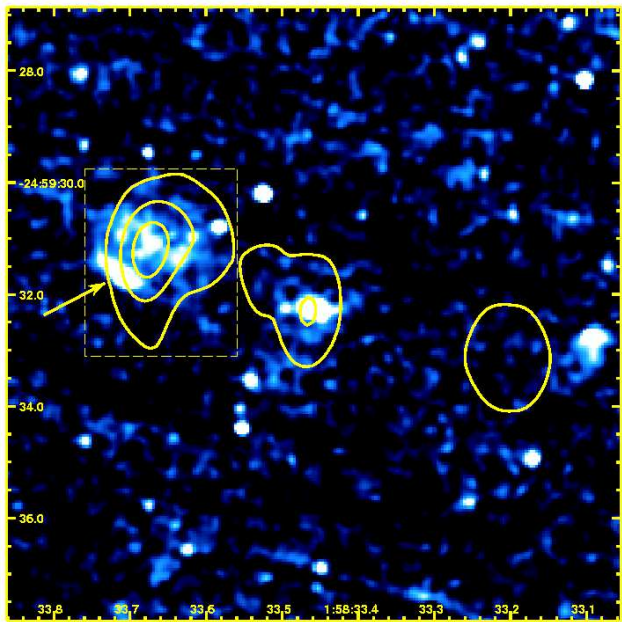}
\caption{\label{fig:0156}HST/WFPC2 F555W image of 0156--252 with 4.7 GHz VLA radio contours superimposed. We have detected extended UV line emission (most likely \ciii) at the position of the NE lobe (box region). An elongated feature in the extended emission resembles a bow-shock that follows the radio lobe (arrow).}
\end{figure*}

Interestingly, the optical flux density of the NE lobe of \si is about 1 order of magnitude higher than predicted. 
We suspect that the optical emission in this lobe is most likely not continuum emission, but line emission. The NE lobe 
coincides with the peak of the Ly$\alpha$ intensity, and several strong emission lines common to HzRG spectra fall within 
the filter. Moreover, the HST/WFPC2 image (Fig. \ref{fig:0156}) shows a bright component roughly following the morphology of the 
radio lobe, reminiscent of a shell of shocked gas. We will treat this feature in a separate discussion below. 

\begin{table*}[t]
\caption[]{\label{tab:synchrotron}Synchrotron models.}
\begin{center}
\begin{tabular}[t]{rccccc|ccc}
\hline
\hline
\multicolumn{1}{c}{Component} & $S_{4.7 GHz}^a$ & $S_{8.2 GHz}^a$ & $\alpha^{8.2}_{4.7}$$^b$ & $S_{X,model}^c$  & $S_{X,obs}^d$ & $\alpha^X_{4.7}$$^e$ & $S_{8.2 GHz,model}^f$ & $S_{5000\AA,model}^g$ \\
                              & (mJy)      &   (mJy)    &                      &    (Jy)      &    (Jy)     &              &    (mJy)&  (Jy)   \\
\hline
\si   NE    & 89 & 47 &   1.15           &  $5.0\times10^{-11}$ &$8.5\times10^{-11}$  & 1.13     & 48     & $1.6\times10^{-7}$ \\
      SW    & 7  & 3.5&   1.25           &  $6.4\times10^{-13}$ &$1.5\times10^{-10}$  & 0.96     & 4.1    & $9.2\times10^{-8}$\\
\hline                                                                                                     
\sii  SE    & 31 & 11 &   1.86           &  $4.8\times10^{-17}$ &$1.2\times10^{-10}$  & 1.05     & 17     & $1.4\times10^{-7}$\\
      NW    & 72 & 32 &   1.46           &  $1.5\times10^{-13}$ &$<1.8\times10^{-10}$  & $>1.07$ &$<40$   & $<2.4\times10^{-7}$ \\
\hline                                                                                                     
\siii NE    & 7  & 2.0&   2.25           &  $1.0\times10^{-20}$ &$<1.4\times10^{-10}$  & $>0.96$ &$<4.1$  & $<8.7\times10^{-8}$ \\
      SW    & 9  & 3.6&   1.65           &  $6.0\times10^{-16}$ &$<1.2\times10^{-10}$  & $>0.98$ &$<5.2$  & $<8.6\times10^{-8}$\\
\hline                                                                                                     
\siv  NE    & 51 & 29 &   1.01           &  $3.7\times10^{-10}$ &$<1.5\times10^{-10}$  & $>1.06$ &$<28$   & $<1.9\times10^{-7}$ \\
      SW    & 37 & 15 &   1.62           &  $4.2\times10^{-15}$ &$7.7\times10^{-11}$  & 1.08     & 20     & $1.1\times10^{-7}$\\
\hline                                                                                                                          
\sv   NE    & 83 & 34 &  1.60            &  $1.4\times10^{-14}$ &$1.3\times10^{-10}$  & 1.10     & 45     & $2.0\times10^{-7}$\\
      SW    & 6  & 1.1&  3.05 (2.2 used) &  $2.1\times10^{-20}$ &$<1.3\times10^{-10}$  & $>0.96$ & $<3.5$ & $<7.7\times10^{-8}$ \\  
\hline
\end{tabular}
\end{center}
$^a$ Integrated radio flux density at 4.7 and 8.2 GHz. The typical error is assumed to be $\sim5\%$.\\
$^b$ Radio spectral index. The typical error is assumed to be $\sim0.2$.\\
$^c$ Predicted flux density at 2 keV for a radio synchrotron model.\\
$^d$ Observed flux density at 2 keV.\\
$^e$ X-ray to (4.7 GHz) radio spectral index.\\
$^f$ Predicted flux density at 8.2 GHz.\\
$^g$ Predicted flux density at 5000\AA.\\
\end{table*}
\begin{table*}[t]
\caption[]{\label{tab:fields}Predictions for SSC and IC/CMB emission, and the equipartition and IC/CMB magnetic field strengths.}
\begin{center}
\begin{tabular}[t]{rlccccrcr}
\hline
\hline
\multicolumn{1}{c}{Component}  & \multicolumn{1}{c}{area$^a$} & $u_s^b$ &  $u_s/u_{CMB}^c$ & $R^d$ & $f_{0.5-6 keV,SSC}^e$ & $\frac{f_{0.5-6 keV,SSC}}{f_{0.5-6 keV,obs}}^f$ & \multicolumn{1}{c}{$B_{eq}^g$}& \multicolumn{1}{c}{$B_{IC/CMB}^h$}  \\
\hline
\si   NE    &  $r=0\farcs35$, $l=1\farcs0$& 2.8 &0.7 &0.04&4.4  &0.4  & 129 & 104\\
      SW    &  $r=0\farcs4$, $l=0\farcs7$ & 0.3 &0.1 &0.02&0.2  &0.01  & 67  & 30\\
\hline                                                             
\sii  SE    &  $r=0\farcs3$, $l=1\farcs2$ & 5.6 &0.9 &0.05&2.7  &0.2  & 169 & 179\\
      NW    &  $r=0\farcs3$, $l=1\farcs2$ & 6.0 &1.0 &0.05&5.1  &$>0.2$  & 170 & $>130$\\
\hline                                                             
\siii NE    &  $r=0\farcs2$, $l=0\farcs7$ & 7.9 &1.2 &0.04&0.7  &$>0.04$  & 213 & $>152$\\
      SW    &  $r=0\farcs2$, $l=0\farcs7$ & 3.3 &0.5 &0.03&0.4  &$>0.03$  & 162 & $>93$\\
\hline                                                             
\siv  NE    &  $r=0\farcs3$              & 2.9 &0.9 &0.04&2.1  &$>0.1$  & 139 & $>44$\\
      SW    &  $r=0\farcs3$, $l=1\farcs0$ & 2.7 &0.8 &0.04&2.0  &0.2  & 135 & 135\\
\hline                                                             
\sv   NE    &  $r=0\farcs5$, $l=1\farcs1$ & 3.4 &0.9 &0.04&6.7  &0.4  & 126 & 154\\
      SW    &  $r=0\farcs6$, $l=2\farcs4$ & 0.3 &0.1 &0.02&0.3  &$>0.02$  & 61 & $>117$\\
\hline
\end{tabular}
\end{center}
$^a$ Cylinder of length $l$ and radius $r$ ($r$ along the line of sight), or a sphere of radius $r$.\\
$^b$ Synchrotron energy density, $u_S$, in units of $10^{-11}$ erg cm$^{-3}$.\\
$^c$ Ratio of the synchrotron to cosmic microwave background energy densities.\\
$^d$ Ratio of the synchrotron to equipartition magnetic field energy densities, $R=u_S/u_B$.\\ 
$^e$ Predicted synchrotron self-Compton (SSC) flux in the 0.5--6 keV band in units of $10^{-16}$ erg s$^{-1}$ cm$^{-2}$.\\
$^f$ Ratio of predicted SSC flux to observed X-ray flux.\\
$^g$ Equipartition magnetic field strength in $\mu G$.\\
$^h$ IC/CMB magnetic field strength in $\mu G$.\\
\end{table*}

\medskip
\noindent
$\bullet$ Synchrotron self-Compton (SSC) emission\\
SSC emission arises due to Inverse-Compton scattering of local 
synchrotron photons. This process 
dominates over other IC mechanisms when the local synchrotron photon energy density is higher than 
that of the external (e.g. CMB) photon field, and has been found to explain the \chandra hot spots 
of radio sources such as 3C 295 \citep{harris00} and Cygnus A \citep{wilson00}. 
We calculate the energy density of the 
local synchrotron photon field, $u_s$, by integrating the radio flux density over 1 decade of frequency (1--10 GHz) and assuming  
a cylindrical geometry for the radio lobes. The results are listed in Table \ref{tab:fields}. 

A detailed calculation of SSC (see e.g. \citet{band85} for the theoretical framework) would require knowledge of the full radio spectrum and the geometry of lobes and hot spots. 
However, we can roughly estimate the SSC flux density in the X-ray from the ratio of the energy losses in the IC and synchrotron channels R=$u_s/u_B\approx L_{IC}/L_S$, 
where $u_B$ is the energy density of the magnetic field $B$, and $L_{IC}$ and $L_S$ are the IC and synchrotron luminosities \citep[see][]{harris00,donahue03}.
We calculate $u_B$ from the magnetic field strength $B_{eq}$ that will make the energy densities of 
fields and particles approximately equal. We use the formula given by \citet{miley80}:
\begin{eqnarray}
B_{eq} & = & 5.69\times 10^{-5}\times\nonumber\\
& &\left[\frac{(1+k)}{\eta}\frac{(1+z)^{3+\alpha_r} S_r \nu_r^{\alpha_r}}{\theta^2~s~\mbox{sin}^{\frac{3}{2}}~\phi}\frac{\nu_2^{\frac{1}{2}-\alpha_r}-\nu_1^{\frac{1}{2}-\alpha_r}}{\frac{1}{2}-\alpha_r}\right]^{\frac{2}{7}},\nonumber
\end{eqnarray}
where $B_{eq}$ is in Gauss, $k$ is the ratio of energy in heavy particles to that in electrons, $\eta$ the filling factor, 
$\theta^2$ (arcsec$^2$) the area taken up by the radio lobe (assumed to be a cylinder viewed broadside), $s$ the pathlength through the lobe (kpc), and 
$\nu_1$ and $\nu_2$ (GHz) are the lower and upper cut-off frequencies of the synchrotron spectrum. We take $\phi=\pi/2$, 
$k=0$ and filling factor unity so as to obtain a real minimum, and note that the results are only weakly dependent on 
these parameters given the 2/7 power. $R$, $B_{eq}$ and $S_{X,SSC}$ are given in Table \ref{tab:fields}, where $B_{eq}\sim100-200$ $\mu$G. In all cases, the ratio of predicted X-ray SSC flux to observed X-ray flux is less than 0.4. 

\medskip
\noindent
$\bullet$ Inverse-Compton (IC) scattering of the CMB\\
Another possibility is the up-scattering of CMB photons by relativistic electrons in the radio source. This effect is 
expected to become significantly more dominant at higher redshifts due to the $(1+z)^4$ increase in the energy density 
of the CMB ($u_{CMB}\sim3.4\times10^{-11}$ erg cm$^{-3}$ at $z\sim2$). 

The scattering electrons are assumed to belong to the same power law distribution of electrons which is responsible for the radio 
synchrotron emission. Therefore, the X-ray flux may be used to constrain the magnetic field strength:
since the energy density of the CMB is fixed at any given $z$, less observed X-ray flux implies fewer electrons, so 
the magnetic field producing the observed radio synchrotron emission must be stronger. 
Using the formalism of \citet{harris79} we calculate the magnetic field strength, $B_{IC}$:

\begin{eqnarray}
B_{IC}^{1+\alpha_r} & =& \left[\frac{(5.05\times10^{4})^{\alpha_r} C(\alpha_r)G(\alpha_r)(1+z)^{3+\alpha_r}S_r\nu_r^{\alpha_r}}{10^{47}S_X\nu_X^{\alpha_r}} \right],\nonumber 
\end{eqnarray}
where $B_{IC}$ is in Gauss, $\alpha_r$ is the radio spectral index, $S_r$ and $S_X$ are the radio and X-ray 
flux densities (erg s$^{-1}$ cm$^{-2}$ Hz$^{-1}$) at $\nu_r$ and $\nu_X$ (both in Hz), respectively.
$C(\alpha_r)$, which is well approximated by the value $1.15\times10^{31}$, and $G(\alpha_r)$, which is 
a function that is slowly varying with $\alpha_r$, can be found in \citet{harris79}. From the radio and 
X-ray flux densities listed in Table \ref{tab:synchrotron} we derive field strengths of $\sim30-180$ $\mu$G for the 
components that are detected in the X-ray. The results are listed in Table \ref{tab:fields}, where we also 
give lower limits on $B_{IC}$ for the undetected lobes.

We can compare $B_{IC}$ to $B_{eq}$, which is an estimate of the magnetic field that is solely based 
on the observed radio synchrotron flux (see expression above). The $B_{eq}$ field strengths are 
typically $\sim100-200$ $\mu$G, remarkably close to the field strengths derived for the IC/CMB mechanism. 
The general agreement that we find between the magnetic field strengths using the two independent 
field estimators is consistent with the X-ray flux being produced by the IC/CMB process. 

The (observed) 1--10 GHZ synchrotron flux in HzRGs is produced by relativistic electrons that 
have $\gamma\sim10^{3.4-3.9}$ for $B\sim100$ $\mu G$ \citep[$\nu_{syn}=4.2(B/1\mu G)\gamma^2$ Hz, see][]{bagchi98}. 
The number density of relativistic electrons with energies between $\gamma$ and $\gamma+d\gamma$ can be expressed as a power-law, 
$N(\gamma)d\gamma=N\gamma^{-s}d\gamma$ with $s=2\alpha+1$. 
The up-scattering of CMB photons to X-ray frequencies is provided by electrons with a Lorentz factor $\gamma\sim10^3$,   
since $\nu_{out}\approx\gamma^2\nu_{in}$, and the frequency for which the energy density of the CMB peaks is $\sim1.6\times10^{11}\times(1+z)$ Hz.
Electrons having such Lorentz factors are highly abundant, given the extrapolation of $N(\gamma)$ to $\gamma\sim10^3$.
Note that in the SSC process described above, the X-ray emission must be produced by the up-scattering of 
synchrotron photons off a population of relativistic electrons having $\gamma\sim10^{4-4.5}$.  
Since $N(\gamma)$ decreases strongly with increasing $\gamma$ and $\alpha>1$ for all our sources, $N(10^{4.5})$ will 
be several orders of magnitude lower 
than $N(10^3)$. Together with the fact that $u_s/u_{CMB}$, is around unity (see Table \ref{tab:fields}) this implies that 
the IC/CMB process will be a far more efficient process for producing X-rays at these redshifts. While the contribution of 
SSC to the observed X-ray flux might still be in the range of $\sim1-40$\% as detailed above, the SSC output is expected to 
peak around $10^{15}$ Hz, or in the UV part of the spectrum.  
  
\medskip
\noindent
$\bullet$ Other mechanisms\\
\noindent
The ambient medium in which a radio source expands is expected to be very different for HzRGs compared to radio sources at 
low redshift. Low redshift sources often lie in a smooth, virialised ($T\sim10^8$ K) atmosphere. For HzRGs such a smooth 
medium does probably not exist because the necessary potential well has not yet formed. Instead they lie in a multiphase medium, 
consisting of cold ($10^4$ K), 
high density clouds embedded in low density regions approaching virial temperatures ($10^{6-7}$). The passage of a radio jet through the multiphase can lead to interesting phenomena, such as jet-induced star formation in the high density clouds and shock heating 
of lower density regions. When intermediate density regions are shocked they may cool off, emitting strong emission lines such as Ly$\alpha$. The lower density gas with high filling factor will be shock heated to temperatures where the gas starts producing X-rays. Can the X-ray emission coincident with the NE lobe of \si be produced by shocks in this thermal gas? 
Interestingly, the NE radio lobe of \si coincides with the peak of the Ly$\alpha$ emission, and the HST/WFPC2 image
(see Fig. \ref{fig:0156}) shows evidence for a shell of shocked (emission-line) gas that follows the bend of 
the radio lobe. 

For 1138--262, \citet{carilli02} propose that much of the
extended X-ray emission comes from ambient gas that is shock heated by the
expanding radio source. For this gas they estimate a density of
0.05 cm$^{-3}$ and a pressure of 10$^{-9}$ dyn cm$^{-2}$.
This pressure is comparable to the optical line emitting gas
and to minimum pressures in the radio source, with a total

gas mass of about $2.5\times10^{12}$ M$_\odot$. Carilli et al.
hypothesize that the high filling factor X-ray emitting gas may
confine both the radio source and the line emitting clouds. 
The extended X-ray luminosity associated with the NE lobe of 0156--252 is a factor 5 or
so less than in 1138--262, implying a factor 5 lower total 
gas mass (for a fixed density) if the X-rays represent shocked gas. 

\begin{figure*}[t]
\includegraphics[width=\columnwidth]{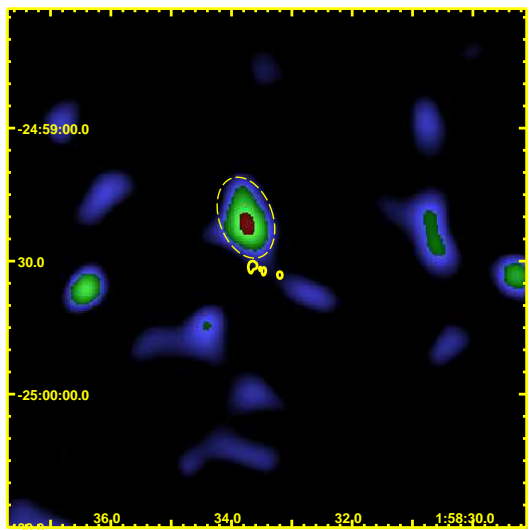}
\caption{\label{fig:egg}X-ray image of the field around \si. The image was smoothed by a 
Gaussian (10\arcsec\ FWHM), after replacing the X-ray counts due to the radio source and 
point sources in the field by a Poissonian background. A large region of diffuse emission 
(indicated by the dashed ellipse) is found just northeast to the radio galaxy (indicated 
by its radio contours).}
\end{figure*}

So far, the discussion has been limited to the X-ray emission directly coincident with the radio lobes and hotspots. 
\citet{brunetti00} describes a mechanism in which IC scattering of (mostly infrared) AGN photons by relativistic 
electrons can also produce X-ray bright emission. Such a mechanism, if it exists, may contribute significantly to the 
radio/X-ray alignment effect. Under the Brunetti mechanism the X-ray luminosity is expected to decrease with increasing 
distance from the source (i.e. the hidden quasar) producing the photons. This will result in opposite gradients in the X-ray and radio luminosities, and entails (in our case) the presence of an inner, undetected part of the radio lobe. 
\citet{brunetti00} predicts that the receding radio lobe will produce brighter X-ray emission than the approaching 
lobe, because time delay makes it closer to the nucleus and because backward scattering is more effective than 
forward scattering. 
In \sii (and possibly in one or several of the other sources as well) there is X-ray emission in between the core and 
the SE lobe. The nature of this emission could be similar to the aligned emission 
seen in several other high-redshift quasars and radio galaxies \citep[e.g.][]{yuan03,fabian03,scharf03} for which the Brunetti  
model is among the possible scenarios. However, the proposed mechanism relies on a (suggested) supply of electrons 
with $\gamma\lesssim100$. Given the highly speculative amplitudes at these low energies, we will not pursue this mechanism 
quantitatively.
 
\subsection{Thermal emission from hot (cluster) gas}

An important driver for observing powerful HzRGs in the X-ray is to search for traces of thermal emission from 
an ICM. To test whether our data show any evidence for the presence of extended, diffuse X-ray emission 
we will attempt two methods:

\medskip
\noindent
$\bullet$ Smoothed fields\\

To obtain upper limits on the luminosity of extended regions in each field, we place 10 circles of 12\arcsec\  
diameter (corresponding to $\sim100$ kpc at $z\sim2$) around each radio source in regions without visible point 
sources. We calculate the average background count in these regions and the $1\sigma$ deviation from the mean. 
We use the resulting $5\sigma$ countrates to calculate $5\sigma$ upper limits on the X-ray flux of typical 100 kpc-sized areas. 
We assumed a Raymond-Smith thermal spectrum with $kT=1$ keV, and used the galactic $n_{HI}$ for each source 
to produce the unabsorbed flux at 0.2--6 keV. We derive an upper limit on the flux of (4--9)$\times10^{-15}$ erg cm$^{-2}$ s$^{-1}$, 
corresponding to luminosities of (1.5--4)$\times10^{44}$ erg s$^{-1}$ (for a uniform sphere).

A $5\sigma$ deviation from the background countrate over a 12\arcsec\ region would easily 
stand out from the X-ray maps after smoothing with a large kernel. Therefore, we create smoothed 
images of diffuse emission. First, we remove counts that are associated with 
the radio galaxy by replacing all counts in a circular region encompassing the entire radio 
structure by the background using the task DMFILTH that maintains the Poissonian nature of the 
background. The background was estimated from a large annular region around the radio source. 
We then run the WAVDETECT algorithm (Freeman et al. 2002) within CIAO to identify the remaining 
point sources and replace them by the background. The images were smoothed by a 10\arcsec\ (FWHM) Gaussian. 
The images were then visually inspected to look for regions of diffuse emission. 
We found a single detection of an extended X-ray component in the smoothed map of \si shown 
in Fig. \ref{fig:egg}. In an ovally shaped region that can be approximated by an ellipse 
of $10\arcsec\times17\arcsec$ ($83\times142$ kpc if it were to be
located at the same distance as the radio galaxy), we find $14.4\pm5$ net counts ($3\sigma$) in the energy 
range (0.2--6) keV with an average energy of 1.2 keV. 
The centroid of the region is situated approximately 12\arcsec\ to the northeast of the core of the 
radio source. The (observed-frame) 0.5--2 keV unabsorbed flux is $2.3\times10^{-15}$ erg cm$^{-2}$ s$^{-1}$.
If this emission were to come from a sphere of thermal gas at the redshift of the radio source, it would correspond to a 
luminosity of $7.4\times10^{43}$ erg s$^{-1}$ and a mass of $10^{11}$ $M_\odot$ (assuming a uniform sphere). 
We could not identify this X-ray region with objects in any of our optical and near-infrared 
imaging data. Therefore, in this particular case the true nature of the emission remains highly 
obscure, and we cannot rule out it being due to noise. However, it illustrates the kind of emission that 
might be detectable in the search for high redshift cluster gas, provided a proper identification can be made.

\begin{figure*}[t]
\begin{minipage}[ht]{\textwidth}
\begin{minipage}[ht]{0.5\textwidth}
\includegraphics[width=\textwidth]{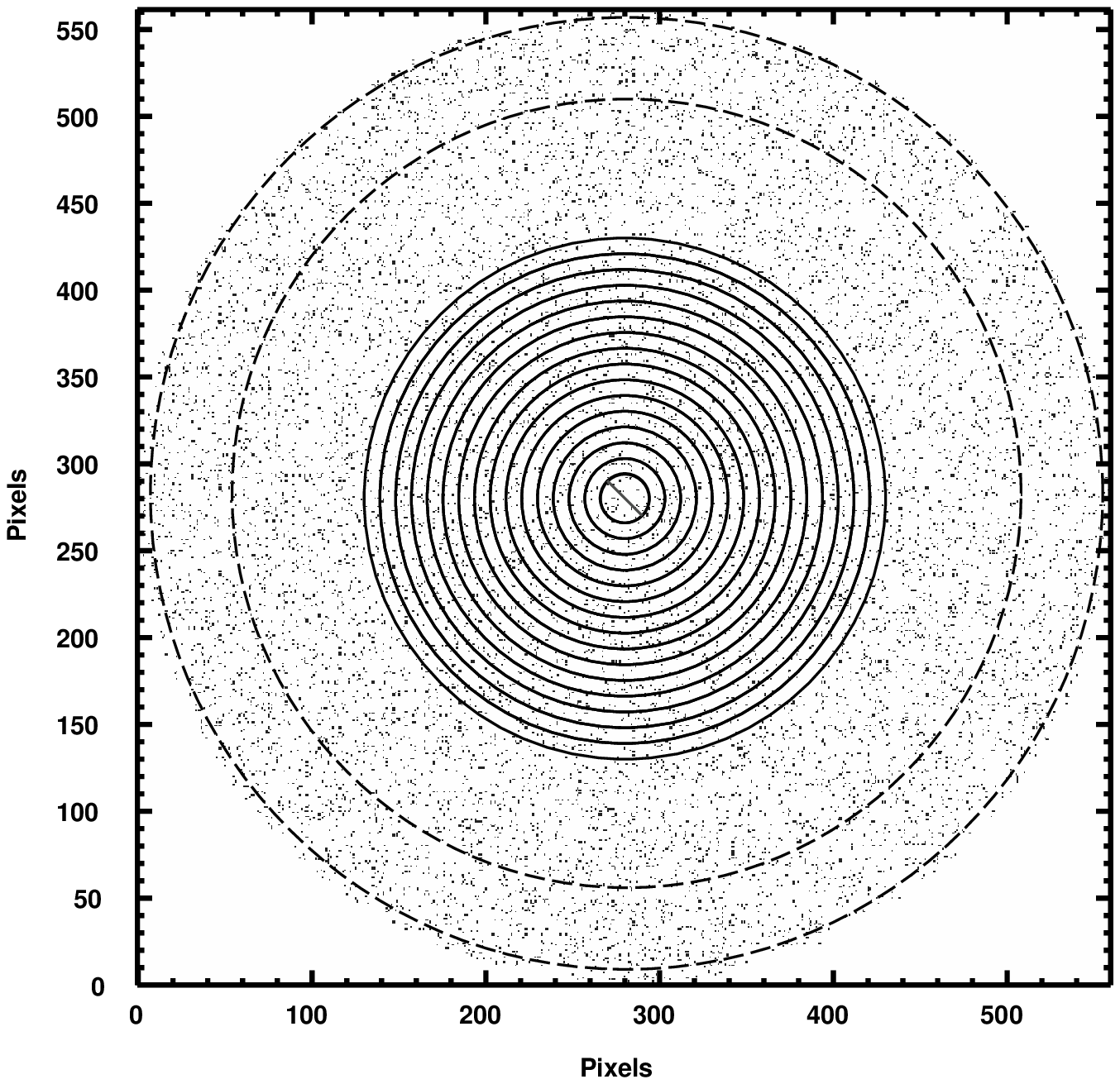}
\end{minipage}
\begin{minipage}[ht]{0.5\textwidth}
\includegraphics[width=\textwidth]{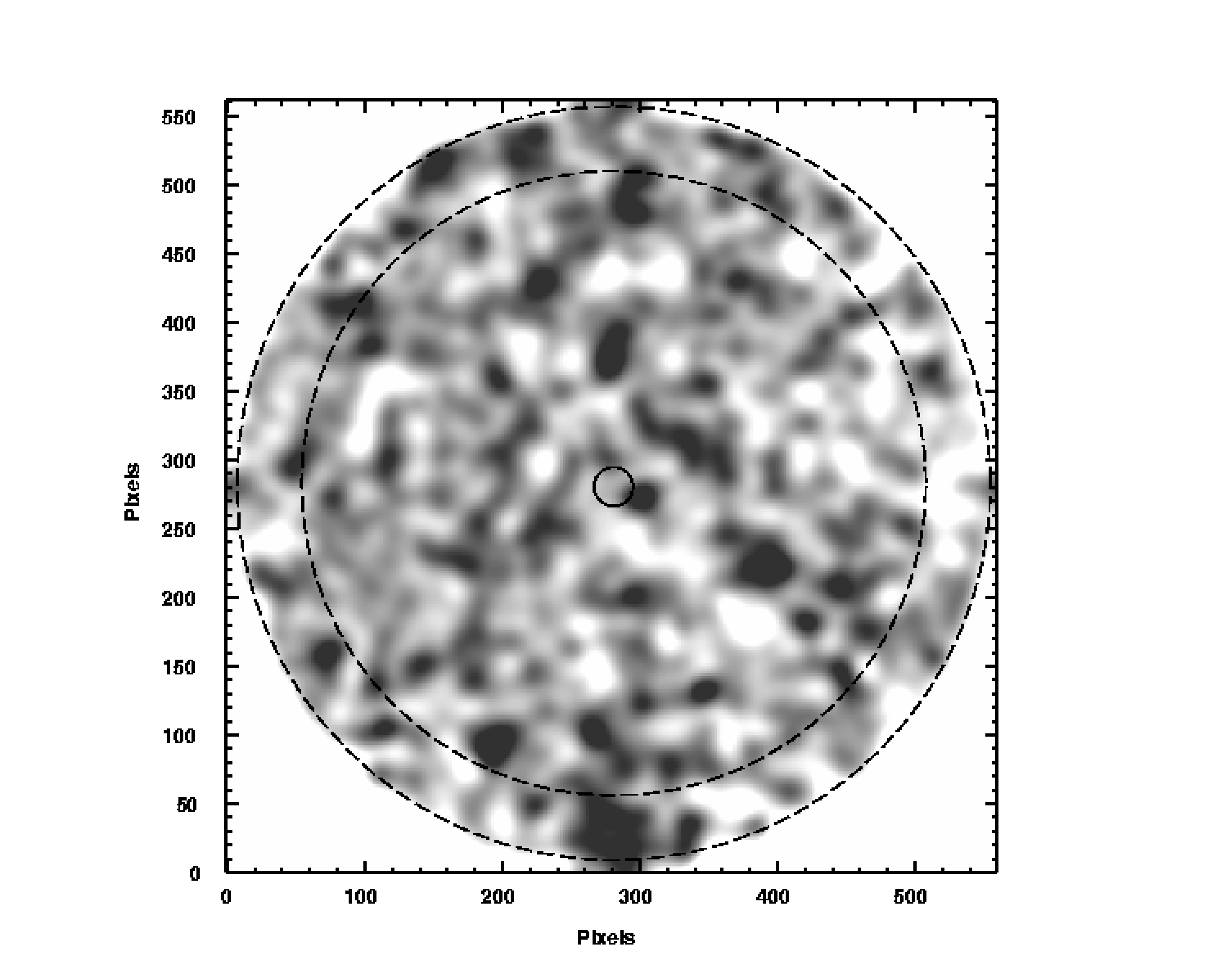}
\end{minipage}
\end{minipage}
\caption{\label{fig:stack}Left: Stacked image of the five HzRG fields, after subtraction of point sources. 
Indicated are the annuli from which a radial surface brightness profile was extracted. The innermost circle encompasses the total extent of the largest radio source in the sample. Counts inside this region due to the radio sources have been replaced by the background, estimated from the large, dashed annular region. Right: Stacked image smoothed by a 10\arcsec (FWHM) Gaussian. Scales run from $2\sigma$ below (white) to $2\sigma$ above (black) the mean. The background region and the maximum size of the radio sources are indicated. The distinct peaks in this summed image are not traced back to significant features in any of the individual images.}
\end{figure*}

\begin{figure*}[t]
\includegraphics[width=\columnwidth]{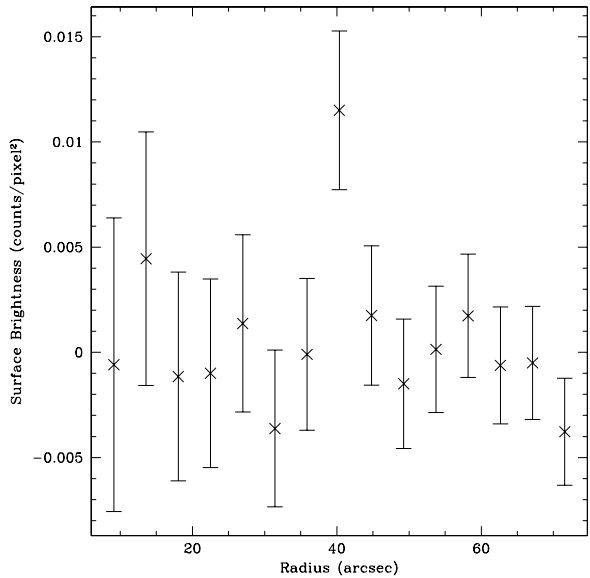}
\caption{\label{fig:rprofile}Radial surface brightness profile extracted from the stacked image shown in Fig. \ref{fig:stack}.}
\end{figure*}

\medskip
\noindent
$\bullet$ Stacked fields\\
To further search for emission from thermal gas in an annular region surrounding the radio sources, 
we stacked the five 20 ks exposures to obtain a single 100 ks exposure of a typical HzRG field. 
Each field was cleared of point sources before stacking them. In the stacked image, we then replaced the entire 
region inside a circle encompassing the largest of the radio 
sources by the background to ensure that we are not measuring extended X-ray emission associated with the radio structures. 
The stacked image was smoothed by a 10\arcsec\ (FWHM) Gaussian. The stacked image and the smoothed image are 
shown in Fig. \ref{fig:stack}. The region used for the background determination is indicated by the large, dashed annulus. 
The small circle in the center of the field indicates the size of the largest radio source. 
From the smoothed image, where 
the scales run from $2\sigma$ below the mean to $2\sigma$ above the mean, we see no evidence for diffuse, extended emission 
in the vicinity of the radio source(s). We obtained a radial profile of surface brightness from the stacked image 
using the annuli as indicated in Fig. \ref{fig:stack}. The radial profile out to 
$1\arcmin$ is consistent with zero contribution from diffuse, extended X-ray emission, as shown in Fig. \ref{fig:rprofile}. 
Out to a radius of $\sim75\arcsec$ ($\sim625$ kpc) we measure $25\pm85$ counts. 
Assuming a 1 keV thermal spectrum and correcting for galactic absorption, the $3\sigma$ upper limit on the (observed) 0.2--6 keV 
flux inside this radius is $1.2\times10^{-14}$ erg cm$^{-2}$ s$^{-1}$, corresponding to a luminosity of 
$<4\times10^{44}$ erg s$^{-1}$. We derive a central electron density of $<7\times10^{-3}$ cm$^{-3}$ and a total enclosed gas mass of 
$<4\times10^{13}M_\odot$, assuming an isothermal sphere with a cluster $\beta$-model surface brightness profile with $\beta=0.67$ and 
a core radius of 200 kpc. 

\subsection{Serendipitous X-ray sources}

\citet{laura02} found evidence of an excess of \chandra detected X-ray sources in the vicinity of 
radio galaxy PKS 1138--262 at $z=2.16$. They found 16 sources in the soft band (0.5--2 keV) with a minimum flux 
of $1\times10^{-15}$ erg cm$^{-2}$ s$^{-1}$, compared to 10.3 (11) calculated from the \chandra Deep Field South (North) for a similarly sized region, and 8 sources with fluxes $\ge3\times10^{-15}$ erg cm$^{-2}$ s$^{-1}$ 
compared to 4 (5.1) for the \chandra Deep Fields. Although this may not be considered a substantial overdensity 
in angular space (an excess of $\sim50\%$ compared to a typical cosmic variance of $20-30\%$), 
it is very significant in redshift space given that six of the serendipitous X-ray sources could be 
identified with previously discovered Ly$\alpha$ or H$\alpha$ emitting galaxies in the protocluster 
surrounding PKS 1138--262 \citep{laura02}. Therefore, observing overdensities of X-ray sources may provide evidence for 
the existence of galaxy overdensities associated with radio sources \citep[see also][]{cappi01}.

Following \citet{laura02}, we use WAVDETECT to 
investigate the presence of serendipitous sources 
in our 5 fields. Because the 2 separate observations of \si\ were taken at different roll angles, 
the corners of the ACIS-S3 chip only have half of the total 20 ks exposure time. We specify a circular region 
 covering the maximum area having the full 20 ks depth. The circle has a radius of 3.852\arcmin, 
thereby covering an area of 46.62\arcmin$^2$. For consistency we use this region in all the fields. We limit the 
energy range for source detection to a maximum of 2 keV \citep[cf.][]{laura02} to improve the $S/N$ 
(the background is lower at lower energies). 
For each field, we exclude the radio source, and we only consider sources that are detected with $S/N>3$, 
where we define $S/N$ as the number of background-subtracted counts divided by the root of the total counts. 
The net count rates of the detected sources were converted into fluxes 
assuming a power-law spectrum with Galactic absorption along the line of sight, and a photon index 
of $\Gamma=2.0$ for the soft band. 
Table \ref{tab:field} lists the number of sources in each field 
that have soft (0.5--2 keV) fluxes of $>1.5$ and $>3\times10^{-15}$ erg cm$^{-2}$ s$^{-1}$. 
Besides the number of sources found in each 46.62\arcmin$^2$ area we give the number density per square degree 
between parentheses. We also list the number of sources in each bin averaged over the 5 fields. For comparison, we have 
run the same detection procedure on the observations of PKS 1138--262 using (i) the full 40 ks exposure,  
and (ii) a 20 ks subsample.  
We have also indicated the number of sources found in the \chandra Deep Fields \citep{mushotsky00,giacconi01}. 
\begin{table*}[t]
\begin{center}
\caption[]{\label{tab:field}The number of X-ray sources detected in the fields of $z\sim2$ radio galaxies, 
and comparison with the field population expected based on the \chandra deep fields.}
\begin{tabular}[t]{lcc}
\hline
\hline
Field           & \multicolumn{2}{c}{Number of sources$^a$}\\
                 & $f^b_{0.5-2 keV}>1.5$ & $f^b_{0.5-2 keV}>3.0$\\
\hline
$0156-252$     & 7 (541) &  5 (386)\\
$0406-244$     & 7 (541) &   7 (541)\\
$0828+193$     & 5 (386) &   4 (309)\\
$2036-254$     & 7 (541) &   5 (386)\\
$2048-272$     & 2 (154) &   1 (77)\\
average of fields & 6 (463) &   4 (309)\\
\hline
$1138-262^{c}$ & 7 (541) &   6 (463)\\
$1138-262^{d}$ & 11 (715) &  7 (541)\\
CDFs$^e$         & 6/7 (470/540) & 3/4 (260/330) \\
\hline
\end{tabular}
\end{center}

\noindent
$^a$ Number of X-ray sources sources found in the field (per deg$^{2}$)\\
$^b$ Minimum source flux in units of $10^{-15}$ erg s$^{-1}$ cm$^{-2}$\\
$^c$ Comparison 1: A 20 ks subsample of the full 40 ks observation of 1138--262\\
$^d$ Comparison 2: The full 40 ks exposure of 1138--262 \citep{laura02}\\
$^e$ Comparison 3: The {\it Chandra} Deep Field South/North \citep{mushotsky00,giacconi01}\\   
\end{table*}

The number of bright sources in the field of \sii is twice as high compared to the \chandra Deep Fields, 
and similarly high as PKS 1138--262. The number of sources in the field of \sv is less than half of 
the number of sources in the CDFs in both flux bins.
On average we find 6 and 4 sources in the $>1.5$ and $>3\times10^{-15}$ erg cm$^{-2}$ s$^{-1}$ flux bins, 
respectively, in good agreement with the CDFs. Based on these number counts of serendipitous sources, 
we find no evidence that the radio sources lie in the same cluster environments as is observed in the case of 
PKS 1138--262 by \citet{laura02}. However, as indicated in Table \ref{tab:field} the factor $\sim2$ 
excess of faint, $>3\sigma$ sources in the field of PKS 1138--262 only becomes apparent in the full 40ks exposure, which may 
indicate that our exposures are not deep enough to make a good comparison.

\begin{figure*}[t]
\includegraphics[width=\columnwidth]{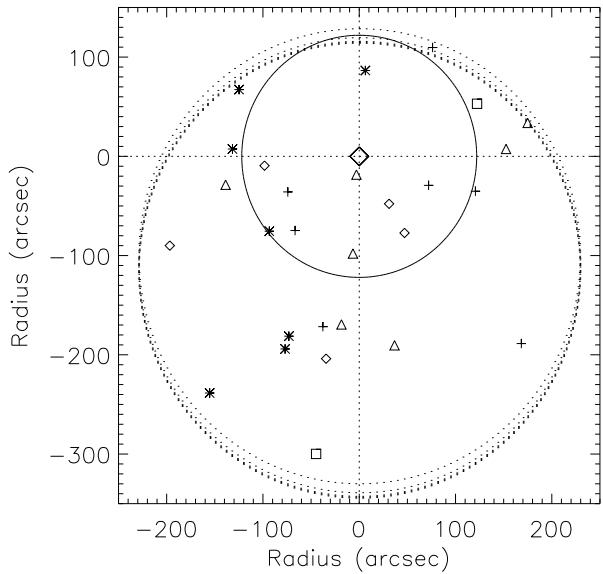}
\caption{\label{fig:agnpositions}The positions of serendipitous X-ray sources with flux $>1.5\times10^{-15}$ erg cm$^{-2}$ s$^{-1}$ in the five HzRG fields combined. Different symbols indicate the sources in the 5 individual fields. The five images are aligned so that the radio sources (indicated by large diamond) all coincide with the position of 0156-252, and rotated to maximize the overlapping areas of the different pointings (indicated by dotted circles). The full circle indicates an area of 1 Mpc in radius around the radio source.}
\end{figure*}

To further study whether there is a preference for field sources to lie in the vicinity of high-redshift radio 
sources, we register the five fields using the radio source positions as centroids, and we rotate the fields 
around their original center so that the area of 
overlap is maximized. This is shown in Fig. \ref{fig:agnpositions}. The number density of sources found 
within 1 Mpc (at $z\sim2$) from the radio sources is 0.77 arcmin$^{-2}$, compared to 0.4 arcmin$^{-2}$ for the density 
of sources outside this region and 0.6 arcmin$^{-2}$ for the average density of the entire field. However, 
the cosmic variance of field X-ray sources is significant \citep[25\%, e.g.][]{cappi01}, making it impossible to conclude 
if serendipitous sources cluster around radio sources on the basis of the current data. Spectroscopy may confirm 
whether some of these sources are associated with the radio galaxies.

\section{Summary and conclusions}
\label{sec:discuss}

We have studied X-ray observations of five radio galaxies at $2<z<2.6$, thereby significantly increasing the number of 
\chandra studies on high redshift radio galaxies. The main conclusions from our analysis are the following. 

The X-ray emission that we detect from the nuclei are consistent with obscured power-law spectra as observed for 
powerful radio galaxies over a wide redshift range \citep[e.g.][]{harris00,carilli02,young02,hardcastle02,scharf03}, 
and can be explained by the unified model in which the broad-line region of radio galaxies is obscured by a dusty 
torus surrounding the nucleus. For 0156--252, this conclusion is confirmed by the fact that 
the best-fit core spectrum is an absorbed power-law with $n(\hbox{H\rm I})\sim2\times10^{22}$ cm$^{-2}$.

Extended X-ray emission coincident with the radio lobes was detected for several of the sources, albeit at low S/N. 
For all but one source, the straight extrapolation from the radio to the X-ray using the radio spectral index rules out 
a synchrotron origin of the emission, unless there are large errors in the radio spectral indices measured between 5 and 8 Ghz. 
Although the predicted X-ray synchrotron flux in source \si is close to the observed value, we have found  
evidence for the X-ray emission being likely associated with shocked, line emitting gas. Our observations confirm that the radio lobes of these high redshift sources may interact with the surrounding (forming) IGM.
 
We interpret the X-ray emission of the remaining sources as being due to the 
inverse Compton scattering of CMB photons off radio synchrotron electrons. This conclusion is supported by the fact that 
our estimates of the IC/CMB and equipartition magnetic field strenghts are in good agreement. Although IC/CMB is 
usually dominated by other processes (e.g. SSC) at low redshift, the relative ease with which it is detected 
in these HzRGs may be ascribed to the $(1+z)^4$ increase in the energy density of the CMB. 

This research was partly motivated by the large RM observed for a significant fraction of HzRGs. Can the large RM be caused by 
cluster-sized atmospheres surrounding the radio sources? Taking reasonable estimates for the pathlength ($\sim100$ kpc)
and the cluster magnetic field strength ($\sim10\mu$G), an intrinsic RM of $\sim1000$ rad m$^{-2}$ would require an ICM 
electron density of $\sim1\times10^{-3}$ cm$^{-3}$. The existence of cores much denser than this is unlikely, since they 
would have an X-ray luminosity of $\gtrsim10^{44}$ erg s$^{-1}$ which has not been observed in any of the HzRGs observed 
to this date. Although higher density material ($n_e\sim200$ cm$^{-3}$) is present in many HzRGs in the form of $10^4$ K 
emission line clouds, the filling factor of this gas ($f_e\sim10^{-6}$) usually implied makes it hard to reproduce the high 
values of RM \citep[e.g.][]{lauraphd}. Alternatively, the large RM can arise from the radio emission passing through a  
sheath of shocked gas surrounding the radio lobes \citep[e.g.][and references therein]{athreya98,carilli02}. The increased gas 
density implied can then explain the large RM if the magnetic field is ordered on scales of only a few kpc. 
If the X-ray emission from the NE lobe of 0156--252 is from shocked gas (as suggested by the bow shock feature seen in 
the HST/WFPC2 image, see Fig. \ref{fig:0156}), then $B\sim13\mu$G given the RM from Table \ref{tab:RM} and the derived 
density of 0.05 cm$^{-3}$.
      
The existence of protoclusters around several HzRGs at $z\sim2$ and higher has been established, mainly through  
 narrow-band, Ly$\alpha$ imaging observations and spectroscopy. However, such studies have not yet been carried out 
for the sources presented in this paper. An overdensity of X-ray sources associated with the well-known galaxy protocluster 
around the radio source 1138--262 at $z=2.2$ was found by \citet{laura02}.
We have analysed the number density of X-ray sources in each of the five \chandra 
fields. None of the fields showed evidence of large-scale structure associated with the radio sources. This may suggest that 
these HzRGs are not found in the same, overdense environments as 1138--262. Similarly, employing several different methods 
we found no evidence for virialised gas, although the upper limits that we derive on thermal gas are not inconsistent 
with a direct, no-evolution extrapolation of local X-ray luminous clusters out to $z\sim2$. Based on 
the X-ray observations presented in this paper alone, the current standing is that these five HzRGs are not in galaxy 
overdensities. However, we remark that X-ray observations are not the most effective way to search for galaxy 
overdensities given the relatively small number fraction of AGN expected at each particular epoch. 

Recent observations of some of the most distant, X-ray luminous clusters made with the {\it Advanced Camera for Surveys} 
aboard HST, show a color-magnitude relation comparable to 
that of local clusters, indicating that early-type galaxies were already well-established by $z\sim1$ \citep{blakeslee03}. 
Thus, the epoch in which the global relations that exist in clusters today are shaped must probably be sought at 
significantly higher redshifts than currently probed (i.e. $z\gtrsim1.3$). 
About a dozen high redshift galaxy overdensities or protoclusters have now been found at $z\gtrsim2$, either 
around HzRGs and quasars or in wide-field surveys.
In the near future we may expect several candidates suitable for a detailed study with \chandra that could determine exactly at what redshift the virialised gas was established. However, the sensitivity required may be of the 
order of that of the \chandra deep fields, due to the extreme cosmological surface brightness dimming. While HzRGs have so 
far presented the best evidence of being associated with massive, forming clusters, they usually come 
with a plethora of non-thermal X-ray mechanisms. Especially if the radio sources lie at the bottom of the 
potential well that coincides with the peak of the bremsstrahlung luminosity, they may not be ideal targets for trying to detect this extemely low surface brightness emission from the ICM.


\begin{acknowledgements}
The work at SAO was partially supported by NASA grant GO2-3139B and contract NAS8-39073. 
We thank Michiel Reuland and Andrew Zirm for productive discussions, and Melanie Johnston-Hollitt for helping out with 
the MIRIAD software. We are grateful to the referee, P. Tozzi, for his many good comments.
\end{acknowledgements}


\bibliographystyle{aa}


\end{document}